\documentclass[11pt]{article}
\usepackage[dvips]{graphicx}

\begin{document}

\begin{center}
\textbf{Reduction of a kinetic model for Na+  channel  activation, and fast and slow inactivation within a neural or cardiac membrane}
 \end{center}
\begin{center} S. R. Vaccaro  \end{center}
\begin{center} 
{\em Department of Physics, University of Adelaide, Adelaide, South Australia, 
5005, 
Australia} \\
  \end{center}
{\em svaccaro@physics.adelaide.edu.au} \\

\begin{quotation}
A fifteen state kinetic model for  Na+ channel gating that describes the coupling between three activation 
sensors, a two-stage fast inactivation process and slow inactivated states, may be reduced to equations
 for a six state system by application of the method of multiple scales. By expressing the occupation
probabilities for closed states and the open state in terms of activation and fast inactivation variables,
and assuming that activation  has a faster relaxation than inactivation and that the activation sensors
are mutually independent, the kinetic equations may be further reduced to rate equations for activation,
and coupled fast and slow inactivation that describe  spike frequency adaptation, a repetitive bursting
oscillation in  the neural membrane, and a cardiac action potential with a plateau oscillation.
The fast inactivation rate function is, in general, dependent on the activation variable $m(t)$ 
but may be approximated by a voltage-dependent function, and the rate function for entry into the
slow inactivated state is dependent on the fast inactivation variable.
\end{quotation}
\newpage

    {\bf INTRODUCTION} 

During prolonged or repetitive depolarization, in addition to the fast inactivation of Na channels that 
contributes to repolarization of the membrane \cite{hh}, a slow inactivation process reduces the number
of Na+ channels available for activation. The increase in slow inactivation of Na+ channels
 during depolarization is associated with a delay to the next spike or a reduction in the firing frequency
 (spike frequency adaptation) \cite{ffg} and is the result of a structural rearrangement in the selectivity
 filter region of the ion channel that generally occurs following the inactivation of the pore
\cite{cmjfc}. Slow inactivation of the transient and persistent components of the Na+ current in a 
mesencephalic V neuron  is associated with the  termination of a bursting oscillation, 
and the increase in the amplitude of the subthreshold oscillation between bursts occurs
during the recovery from slow inactivation \cite{wc}. In subicular neurons adjacent  to the 
hippocampus,  the transition from bursting to single spiking is influenced by the slow inactivation
of Na+ channels, and this may provide a mechanism for enhancing the effect of input signals \cite{ccs}.

For Na+ channels with slow inactivation, the Na+ current $I_{Na}$ may be described by the expression 
$ m^3 hs(V_{Na} - V)$ \cite{ffg} where $V_{Na}$ is the equilibrium potential, and 
the activation variable $m$, the fast inactivation variable $h$, 
 and the slow inactivation variable $s$ satisfy the equations
\begin{eqnarray}
\frac{dm}{dt}       & = &   \alpha_m - m(t)(\alpha_m + \beta_m)   \label{mmm} \\
\frac{dh}{dt}       & = &   \alpha_h - h(t)(\alpha_h + \beta_h)   \label{hhh} \\
\frac{ds}{dt}       & = &   \alpha_s - s(t)(\alpha_s + \beta_s)   \label{sss} 
\end{eqnarray}
and the rate functions $\alpha_g$ and $\beta_g$ are dependent on the membrane potential $V$ for $g = m,h$
and $s$. The  Na+ current may also be expressed as $O(t)(V_{Na}-V)$ where $O(t)$ is the open
state probability that is determined by a kinetic model   where transitions between states represent
the activation of three voltage sensors to open the channel, a two stage fast inactivation process
\cite{cgabc} and subsequent slow inactivation \cite{cr}.
This model is consistent with a recent study, based on  the effect of molecular inhibitors
on Na+ channel gating, that has proposed that fast and slow Na+  channel inactivation
are sequential processes  \cite{osijb} and that the activation of the DIV sensor has 
an essential role in each type of inactivation  \cite{mgh}.

Single channel recording techniques have demonstrated that ion channels are thermally activated between
 closed and open states \cite{ns}, and therefore,  the Hodgkin Huxley (HH) equations describe  the behavior
 of a large number  of stochastic Na+ and K+  channels. The probability distribution 
for the number N of open Na+ channels satisfies a  master equation, and 
for sufficiently large N, by application of a system size expansion, the master equation  may be  
approximated by a Fokker-Planck equation \cite{fl}. As the diffusion terms are small, it may be further reduced to
deterministic equations that are equivalent  to the rate equations for the activation variable m and the 
inactivation variable h. 

Assuming that each voltage sensor is a Brownian particle in an energy landscape, the master equation 
for the random walk within the membrane may be reduced to a Smoluchowski equation that is dependent
on a diffusion parameter and a potential of mean force \cite{vank}. As the relaxation within each
deep well is rapid, the probability density may be expressed as the product of the stationary distribution 
and a survival probability that is the solution of a rate equation \cite{gard}. By approximating
the potential function for the voltage sensor by a square well potential, the low frequency component 
of the solution of the Smoluchowski equation may be expressed as  differential equations for the 
survival probabilities of the closed  and open states \cite{vac2,vac3}, and  is  
similar to that obtained from a numerical solution \cite{sqb}.

For  a system of differential equations that has a separation of time scales,
a reduced system may be  derived explicitly by expressing the solution as an
asymptotic expansion that is dependent on the fast and slow time scales \cite{kc}.
 A  variable that attains a quasi-steady state after an initial fast transient, is the
solution of  an approximate algebraic equation that may be obtained as the lowest order term in an
asymptotic expansion of the solution of the full system, and therefore, the long-time 
behavior is governed by the dynamics of the slow variables that form a subsystem of lower dimension.
The method of multiple scales and other singular perturbation techniques have been applied to
the equations in many areas of physics and biology such as orbital mechanics, 
coupled nonlinear oscillators and biochemical and enzyme reactions \cite{kc,ks}.

In this paper, it is shown that  by taking account of the large relative  magnitude of the transition
rates between some states,  a fifteen state kinetic model that describes Na+ channel gating with
three activation sensors, a two-stage fast inactivation process, and a slow transition to  additional
inactivated states, may be approximated by  equations for a six state system. Assuming that the
activation sensors are mutually independent and activation  has a smaller relaxation time than 
fast inactivation, the inactivation rate function is, in general, dependent on the activation variable
$m(t)$ but may be approximated by a voltage-dependent function, and the
slow inactivation rate function is dependent on the fast inactivation variable $h_f(t)$. 
The kinetic model  describing Na+ channel gating may be further reduced to rate equations
for activation, and fast and slow inactivation with a solution that may exhibit spike frequency adaptation,
a repetitive bursting oscillation and a cardiac action potential with  a plateau oscillation.

  {\bf REDUCTION OF A KINETIC MODEL FOR Na+ CHANNEL ACTIVATION  AND FAST   INACTIVATION }
 
By assuming that Na+ channel activation and inactivation are  independent, the Hodgkin-Huxley (HH)  
rate equations for Na+ and K+ channels and the membrane current equation provide a good account of the action
potential waveform, the threshold potential and subthreshold oscillations in the squid axon membrane
\cite{hh}, and the approach has been applied to a wide range of voltage-dependent ion channels in nerve, muscle and
cardiac membranes  \cite{hille}. However, subsequent experimental studies have shown that the probability
of Na+ channel inactivation increases with the degree of  activation of the channel \cite{ab}, the recovery 
from inactivation is more probable following  deactivation \cite{kb}, and the kinetic equations for 
coupled Na+  activation and inactivation  processes describe ion channel states and their transitions,
and provide a good description of  the ionic and gating currents during a voltage clamp \cite{cgabc}.

If the Na+ channel  conductance is dependent on the activation of three voltage sensors  coupled to 
a two-stage inactivation process \cite{cgabc}, the kinetics may be described by a twelve state kinetic model
 (see Fig. 1) where the occupation probabilities of the closed states $C_1$, $C_2$, $C_3$, $A_1$, $A_2$
 and $A_3$, the open states $O$ and $A_4$ and the inactivated states $I_1$, $I_2$, $I_3$ and $I_4$ are 
determined by the equations
\begin{eqnarray}
\frac{dC_1}{dt} & = &  -(\alpha_{i1} + \alpha_{C1})C_1(t)  + \beta_{C1} C_2(t) + \beta_{i1} A_1(t) \label{12c1} \\
\frac{dC_2}{dt} & = &  -(\alpha_{i2} + \alpha_{C2} + \beta_{C1})C_2(t)  + \alpha_{C1} C_1(t)  \nonumber  \\
                &   &  +\beta_{C2} C_3(t) + \beta_{i2} A_2(t) \label{12c2} \\
\frac{dC_3}{dt} & = &  -(\alpha_{i3} + \alpha_O + \beta_{C2})C_3(t)  + \alpha_{C2} C_2(t)  \nonumber  \\
                &   &  +\beta_O O(t) + \beta_{i3} A_3(t) \label{12c3} \\
\frac{dO}{dt} & = &  - (\beta_O + \alpha_{i4})O(t) + \alpha_O C_3(t)  + \beta_{i4} A_4(t) \label{12o} \\
\frac{dA_1}{dt} & = &   -(\alpha_{A1} + \beta_{i1} + \gamma_{i1})A_1(t) + \alpha_{i1} C_1(t) \nonumber \\
               &   & + \delta_{i1} I_1(t) + \beta_{A1} A_2(t) \label{12a1} \\
\frac{dA_2}{dt} & = &  -(\alpha_{A2} + \beta_{A1} + \beta_{i2}  + \gamma_{i2})A_2(t)  \nonumber \\
                &   & + \alpha_{i2} C_2(t) + \delta_{i2}I_2(t) + \alpha_{A1} A_1(t) + \beta_{A2} A_3(t) \label{12a2} \\
\frac{dA_3}{dt} & = &  -(\alpha_{A3} + \beta_{A2} + \beta_{i3}  + \gamma_{i3})A_3(t)  \nonumber \\
                &   & + \alpha_{i3} C_3(t) + \delta_{i3}I_3(t) + \alpha_{A2} A_2(t)+ \beta_{A3} A_4(t)  \label{12a3} \\
\frac{dA_4}{dt} & = &  -(\beta_{A3} + \beta_{i4}  + \gamma_{i4})A_4(t)  \nonumber \\
                &   &  + \alpha_{i4} O(t) + \delta_{i4}I_4(t) + \alpha_{A3} A_3(t)  \label{12a4} \\
\frac{dI_1}{dt} & = & -(\alpha_{I1} + \delta_{i1}) I_1(t) + \gamma_{i1} A_1(t) + \beta_{I1} I_2(t) \label{12i1} \\
\frac{dI_2}{dt} & = &  - (\alpha_{I2} + \beta_{I1} + \delta_{i2}) I_2(t) \nonumber  \\
                &   &  +\gamma_{i2} A_2(t) +\alpha_{I1} I_1(t) +\beta_{I2} I_3(t), \label{12i2} \\
\frac{dI_3}{dt} & = &  - (\alpha_{I3} + \beta_{I2} + \delta_{i3}) I_3(t) \nonumber  \\
		&   &  +\gamma_{i3} A_3(t) +\alpha_{I2} I_2(t)+ \beta_{I3} I_4(t) . \label{12i3} \\
\frac{dI_4}{dt} & = & - ( \beta_{I3} + \delta_{i4}) I_4(t) + \gamma_{i4} A_4(t) +\alpha_{I3} I_3(t)  . \label{12i4}
\end{eqnarray}

It is assumed that Na+ channels depolarize the membrane, K+ and leakage channels repolarize the membrane,
and the K+ conductance is proportional to $n(t)^4$ where the activation variable    $n(t)$   satisfies
the equation \cite{hh}
\begin{equation}
\frac{dn}{dt}   =  \alpha_n  - n(t)(\alpha_n + \beta_n), \label{nn} 
\end{equation}
and $\alpha_n$ and $\beta_n$ are voltage dependent rate functions. This equation may be derived from a 
kinetic model for K+ channel gating where the voltage dependence of $\alpha_n$ and $\beta_n$ may be
 expressed in terms of the transition rates for a two stage voltage sensor activation  process  \cite{vac4,zha}.  Therefore, the membrane current equation is 
\begin{equation}
C\frac{dV}{dt}= i_e - \bar{g}_{Na} O(t) (V - V_{Na}) - \bar{g}_{K} n(t)^4 (V - V_{K}) - 
\bar{g}_{L} (V - V_{L}),  \label{cur3}
\end{equation}
where $\bar{g}_{j}$ is the  conductance, $V_{j}$ is the equilibrium potential
 for each channel $j$ (Na+, K+ and  leakage), and $i_e$ is the external current.

When the fast inactivation transition rates $\alpha_{ik}   \ll \gamma_{ik}$,
 $ \delta_{ik}  \ll  \beta_{ik}$, and $\gamma_{ik} + \beta_{ik}$
is greater than the activation and deactivation rate functions, for each $k$,
the occupation probabilities of $A_1$ to $A_4$ attain quasi-stationary values in a time that
is smaller than the relaxation of the membrane potential and the  closed, open and inactivated states
\cite{vac1}, and Eqs. (\ref{12c1}) to  (\ref{12i4})  may be reduced to an eight state system by 
expressing the solution as a two-scale asymptotic expansion and eliminating secular terms \cite{kc}
(see Fig. 2)
\begin{eqnarray}
\frac{dC_1}{dt} & = &  -(\rho_1 + \alpha_{C1})C_1(t) + \beta_{C1} C_2(t) + \sigma_1 I_1(t) \label{8c1} \\
\frac{dC_2}{dt} & = &   - (\alpha_{C2} + \beta_{C1} + \rho_2) C_2(t) + \alpha_{C1} C_1(t) + \beta_{C2} C_3(t) + \sigma_2 I_2(t) \label{8c2} \\
\frac{dC_3}{dt} & = &   - (\alpha_{O} + \beta_{C2} + \rho_3) C_3(t) + \alpha_{C2} C_2(t) + \beta_{O} O(t) + \sigma_3 I_3(t) \label{8c3} \\
\frac{dO}{dt}   & = &    - (\beta_O + \rho_4) O(t) + \alpha_O C_3(t) + \sigma_4 I_4(t)  \label{8c4} \\
\frac{dI_1}{dt} & = &   -(\alpha_{I1} + \sigma_1) I_1(t) + \rho_1 C_1(t) + \beta_{I1} I_2(t)  \label{8c5} \\
\frac{dI_2}{dt} & = &    -(\alpha_{I2} +\beta_{I1}  + \sigma_2) I_2(t) + \alpha_{I1}I_1(t) + \beta_{I2} I_3(t) 
 + \rho_2 C_2(t) \label{8c6} \\
\frac{dI_3}{dt} & = &     -(\alpha_{I3} +\beta_{I2} + \sigma_3) I_3(t) + \alpha_{I2}  I_2(t) + \beta_{I3} I_4(t) + \nonumber  \\
                      &    &    \rho_3 C_3(t),  \label{8c7} \\
\frac{dI_4}{dt} & = &     - (\beta_{I3} + \sigma_4) I_4(t)  + \alpha_{I3} I_3(t)  + \rho_4 O(t),  \label{8c8} 
\end{eqnarray}
where the derived rate functions for Na+ channel inactivation and recovery are, for each $k$,
\begin{eqnarray}
\rho_k    & = &  \frac{\alpha_{ik}\gamma_{ik}}{\beta_{ik} + \gamma_{ik}},  \label{rhok}  \\
\sigma_k  & = & \frac{\delta_{ik} \beta_{ik}  }{\beta_{ik} + \gamma_{ik}}  \label{sigk}.
\end{eqnarray}
The Na+ and K+ channel activation rate functions between closed and open states may also be expressed 
in terms of  the transition rates of a two or three stage process \cite{vac4}, that are dependent 
on electrostatic and hydrophobic forces on the charged residues of the S4 voltage sensor \cite{llg}.

If it is assumed that the inactivation sensor and the three activation sensors are independent,
the HH rate equations for  Na+ channel  activation and inactivation are exact solutions of an eight 
state kinetic model for  channel gating \cite{hille, keener}. However, activation and inactivation are 
coupled processes, and if $\alpha_{I1} \gg \rho_1 $  and  $\sigma_1 \gg \beta_{I1}$, based on 
empirical rate functions for a Na+ channel  \cite{cgabc}, by expressing the solution as an
asymptotic expansion that is dependent on fast and slow time scales and solving the equations to lowest 
order \cite{kc},  Eqs.  (\ref{8c1}) and (\ref{8c6}) may be approximated  by  (see Fig. 3) 
\begin{eqnarray}
\frac{dC_1}{dt} & = &  -(\alpha_{C1} + \hat{\rho}_1) C_1(t) + \beta_{C1} C_2(t) + \hat{\sigma}_1 I_2(t)
 \label{8c1a} \\
\frac{dI_2}{dt} & = &  - (\alpha_{I2} + \hat{\sigma}_1 + \sigma_2) I_2(t)
 + \beta_{I2} I_3(t) + \hat{\rho}_1  C_1(t) + \rho_2 C_2(t) \label{8c6a} 
\end{eqnarray}
where  
\begin{eqnarray}
\hat{\rho}_1    & = &  \frac{\rho_1 \alpha_{I1}}{\alpha_{I1} + \sigma_1},  \label{rho1}  \\
\hat{\sigma}_1  & = & \frac{\sigma_1 \beta_{I1}  }{\alpha_{I1} + \sigma_1},  \label{sig1}
\end{eqnarray}   
\begin{equation}	
I_1(t) \approx   \frac{ \rho_1 C_1(t) +  \beta_{I1} I_2(t)}{\alpha_{I1} + \sigma_1} \label{8c5a} 
\end{equation}
and n and V are determined by  Eqs. (\ref{nn}) and  (\ref{cur3})  (see Fig. 4).

In Eqs. (\ref{8c7}), (\ref{8c8})  and  (\ref{8c6a}), it is assumed that
the transition rates between fast inactivated states
with occupation probabilities  $I_2$,  $I_3$ and $I_4$  are an order of magnitude larger than
inactivation  and recovery rates, and activation and deactivation rates between closed and open states,
  and therefore, by expressing the solution as a two-scale asymptotic expansion and
eliminating secular terms \cite{kc},  it may be shown  that Eqs. (\ref{8c1}) to (\ref{8c8}) may be reduced
to a five state kinetic model (see Fig. 5)  
\begin{eqnarray}
\frac{dC_1}{dt} & = &  -(\rho_1 + \alpha_{C1})C_1(t) + \beta_{C1} C_2(t) +\hat{\sigma}_{1r}  I(t)     \label{is1} \\
\frac{dC_2}{dt} & = &   - (\alpha_{C2} + \beta_{C1} + \rho_2) C_2(t) + \alpha_{C1} C_1(t) + \beta_{C2} C_3(t) + \sigma_{2r}  I(t) \label{is2} \\
\frac{dC_3}{dt} & = &  - (\alpha_{O} + \beta_{C2} + \rho_3) C_3(t) +  \alpha_{C2} C_2(t) + \beta_{O} O(t)  + \sigma_{3r}  I(t)  \label{is3} \\
\frac{dO}{dt}    & = &   - (\beta_O + \rho_4) O(t) +  \alpha_O C_3(t) + \sigma_{4r} I(t)  \label{sis4} \\
\frac{dI}{dt}    & = &   -(\hat{\sigma}_{1r} + \sigma_{2r} +  \sigma_{3r} +  \sigma_{4r}) I(t) +  \nonumber  \\
                      &    &  \hat{\rho}_1 C_1(t) + \rho_2 C_2(t) + \rho_3 C_3(t) + \rho_4 O(t)  \label{is5} 
\end{eqnarray}
where $C_1(t) + C_2(t) + C_3(t)+O(t) + I(t) = 1$ and
\begin{eqnarray}
\hat{\sigma}_{1r} & =  &  \frac{\hat{\sigma}_1  \beta_{I2} \beta_{I3}}{\alpha_{I2}\alpha_{I3} + \alpha_{I2}\beta_{I3} +\beta_{I2}\beta_{I3}}  \label{sig1r} \\
\sigma_{2r}       & =  &  \frac{\sigma_2 \beta_{I2} \beta_{I3}  }{\alpha_{I2}\alpha_{I3} + \alpha_{I2}\beta_{I3} +\beta_{I2}\beta_{I3}}   \label{sig2r} \\
\sigma_{3r}       & =  &  \frac{\sigma_3   \alpha_{I2} \beta_{I3}}{\alpha_{I2}\alpha_{I3} + \alpha_{I2}\beta_{I3} +\beta_{I2}\beta_{I3}} \label{sig3r} \\
\sigma_{4r}       & =  &  \frac{\sigma_4   \alpha_{I2} \alpha_{I3}}{\alpha_{I2}\alpha_{I3} + \alpha_{I2}\beta_{I3} +\beta_{I2}\beta_{I3}} \label{sig4r}.
\end{eqnarray}
Following an initial transient, it may be shown that $I_2(t)$, $I_3(t)$ and $I_4(t)$ may be approximated by
\begin{eqnarray}
I_2(t)    & \approx &  \frac{\beta_{I2} \beta_{I3} I(t) }{\alpha_{I2}\alpha_{I3} + \alpha_{I2}\beta_{I3} +\beta_{I2} \beta_{I3}}    \label{i2} \\
I_3(t)   & \approx & \frac{\alpha_{I2} \beta_{I3} I(t) }{\alpha_{I2}\alpha_{I3} + \alpha_{I2}\beta_{I3} +\beta_{I2} \beta_{I3}}  \label{i3} \\
I_4(t)   & \approx & \frac{\alpha_{I2} \alpha_{I3} I(t) }{\alpha_{I2}\alpha_{I3} + \alpha_{I2}\beta_{I3} +\beta_{I2} \beta_{I3}},  \label{i4}
\end{eqnarray}
where $I(t) = I_2(t) + I_3(t) + I_4(t)$. Eqs. (\ref{i2}) to (\ref{i4}) may also be obtained
by application of singular perturbation analysis to a kinetic model for a cardiac Na+ channel  \cite{sb}. 
During an action potential, the solution of  Eqs. (\ref{8c1}) to (\ref{8c8}) may be approximated  by
 the solution of  Eqs. (\ref{is1}) to  (\ref{is5}), where n and V are determined by  Eqs.  (\ref{nn})  
and  (\ref{cur3}), and $I_1$ to $I_4$ are calculated from Eqs. (\ref{8c5a}) and  
(\ref{i2}) to (\ref{i4})  (see Fig. 6).

Assuming that $C_1(t) = m_1(t) h(t)$, $C_2(t) = m_2(t) h(t)$, $C_3(t) = m_3(t) h(t)$, $O(t) = m_O(t) h(t)$ and
$I(t) = 1 - h(t)$, where $m_1(t),m_2(t),m_3(t) $ and $m_O(t)$ are activation variables and $h(t)$ is an inactivation variable,
 Eqs. (\ref{is1}) to  (\ref{is5}) may be expressed as
\begin{eqnarray}
\frac{dm_1}{dt} & = &  -(\rho_1 + \alpha_{C1} + \sigma(t) - \rho(t))m_1(t) + \beta_{C1} m_2(t) +  \nonumber  \\
                      &    &  \hat{\sigma}_{1r} (1/h(t) - 1)      \label{sm1} \\
\frac{dm_2}{dt} & = &   - (\alpha_{C2} + \beta_{C1} + \rho_2 + \sigma(t) - \rho(t)) m_2(t) + \alpha_{C1} m_1(t) + \beta_{C2} m_3(t) + \nonumber  \\
                      &    & \sigma_{2r}  (1/h(t) - 1) \label{sm2} \\
\frac{dm_3}{dt} & = &   - (\alpha_{O} + \beta_{C2} + \rho_3 + \sigma(t) - \rho(t)) m_3(t) +\alpha_{C2} m_2(t) + \beta_{O} m_O(t)  + \nonumber  \\
                      &    &  \sigma_{3r}  (1/h(t) - 1) \label{sm3} \\
\frac{dm_O}{dt}   & = &    - (\beta_O + \rho_4 + \sigma(t) - \rho(t)) m_O(t) + \alpha_O m_3(t) +  \nonumber  \\
                      &    &  \sigma_{4r} (1/h(t) - 1)  \label{sm4} \\
\frac{dh}{dt} & = &  \hat{\sigma}_{1r} + \sigma_{2r} +  \sigma_{3r} +  \sigma_{4r} -  \nonumber  \\
                  &    & h(t)(\hat{\sigma}_{1r} + \sigma_{2r} +  \sigma_{3r} +  \sigma_{4r} + \rho(t) )   \label{sm5} 
\end{eqnarray}
where 
\begin{eqnarray}
\rho(t)  & = &  \hat{\rho}_1 m_1(t) + \rho_2 m_2(t) + \rho_3 m_3(t) +  \rho_4 m_O(t)  \label{rhosum2} \\
 \sigma(t) & = & ( \hat{\sigma}_{1r}  + \sigma_{2r}  + \sigma_{3r} + \sigma_{4r}) (1/h(t) - 1).  \label{sigmasum2} 
\end{eqnarray}
The inactivation  rates $\rho_k$ and recovery rates $\sigma_{k}$, for each $k$, are an order of magnitude
smaller than the activation and deactivation rates, and therefore, from an asymptotic expansion of the 
solution, it may be shown to lowest order that Eqs. (\ref{sm1}) to (\ref{sm4}) for the activation variables 
may be approximated by (see Fig. 7)
\begin{eqnarray}
\frac{dm_1}{dt}   & = &  -\alpha_{C1} m_1(t) + \beta_{C1} m_2(t)      \label{sm1a} \\
\frac{dm_2}{dt}   & = &  - (\alpha_{C2} + \beta_{C1}) m_2(t) + \alpha_{C1} m_1(t) + \beta_{C2} m_3(t)  \label{sm2a} \\
\frac{dm_3}{dt}   & = &   - (\alpha_{O} + \beta_{C2}) m_3(t) + \alpha_{C2} m_2(t) + \beta_O m_O(t)  \label{sm3a} \\
\frac{dm_O}{dt}   & = &    - \beta_O m_O(t) + \alpha_O m_3(t).  \label{sm4a} 
\end{eqnarray}
That is, the inactivation and recovery rates, and the variable $h(t)$, generally only have a small effect
on the time-dependence of the activation variables.

If the activation sensors are mutually independent ($\alpha_{C1} = 3 \alpha_m, \alpha_{C2} = 2 \alpha_m, 
\alpha_O =  \alpha_m, \beta_{C1} =  \beta_m,  \beta_{C2} = 2 \beta_m,\beta_{O} = 3 \beta_m $), 
 Eqs. (\ref{sm1a}) to (\ref{sm4a}) have the solution $m_1(t) = (1 - m(t))^3 $, $m_2(t) =  3 m(t)(1 - m(t))^2$,
 $m_3(t) =  3 m(t)^2(1 - m(t))$, $m_O(t) = m(t)^3 $,  where $m(t)$ satisfies
\begin{equation}
\frac{dm}{dt}   =  \alpha_m  - m(t)(\alpha_m + \beta_m), \label{mm3} 
\end{equation}
and therefore, from Eq. (\ref{rhosum2}), 
\begin{eqnarray}
\rho(t)  & = &  \hat{\rho}_1 (1 - m(t))^3 + 3 \rho_2 m(t)(1 - m(t))^2 + 3 \rho_3 m(t)^2(1 - m(t)) + \nonumber  \\
        &    &  \rho_4 m(t)^3.  \label{rhosum3} 
\end{eqnarray}
However,  as the activation variable generally has a faster time constant than $h(t)$,  
 $\rho(t)$ may be approximated by 
\begin{equation}
\beta_h = \hat{\rho}_1 (1 - m_\infty)^3 + 3\rho_2 m_\infty(1 - m_\infty)^2 + 3\rho_3  m_\infty^2(1 - m_\infty)+
  \rho_4 m_{\infty}^3 \label{beth4} 
\end{equation}
where $ m_\infty =  \alpha_m /(\alpha_m + \beta_m)$  for each membrane potential, and
$\beta_h$ is a voltage dependent function, as assumed by HH \cite{hh}. The activation function $m_\infty$
and each inactivation rate $\rho_k$ has an exponential voltage dependence for a small depolarization but
for larger potentials,  the variation has a  plateau, and therefore, accounts for the voltage dependence of 
$\beta_h$ (see Fig. 8). 

Eq. (\ref{sm5}) may be expressed as 
\begin{equation}
\frac{dh}{dt}        =   \alpha_h - h(t)(\alpha_h + \beta_h)   \label{sm5a} 
\end{equation}
where
\begin{equation}
\alpha_h = \hat{\sigma}_{1r} + \sigma_{2r} +  \sigma_{3r} +  \sigma_{4r},  \label{alfh4} 
\end{equation}
and as $\sigma_{2r}, \sigma_{3r}, \sigma_{4r} \ll  \hat{\sigma}_{1r}$, $ \alpha_h \approx \hat{\sigma}_{1r}$. 
For a moderate hyperpolarization ($\sigma_1  \gg \alpha_{I1}, \beta_{I1}$),
$ \hat{\sigma}_{1r}   \approx \hat{\sigma}_{1} \approx  \beta_{I1}$, and therefore, the voltage dependence 
of  $ \alpha_h$ is approximately exponential \cite{hh} (see Fig. 8), but may attain a plateau value for a
 large hyperpolarization  \cite{cgabc,kb,vac1}.

If the previous conditions for each stage of reduction are satisfied,  the solution of the
twelve state kinetic model, Eqs. (\ref{12c1}) to (\ref{12i4}), during an action potential,
may be approximated by the solution of  Eqs. (\ref{mm3}) and (\ref{sm5a}), where $n$ and $V$ are
determined by Eqs.  (\ref{nn})  and  (\ref{cur3})   - see Fig. 9
for a  Na+ channel with an inactivation rate independent of the closed or  open  state
\cite{hh}, and Fig. 10 for a channel where the Na+ inactivation rate  increases with
the degree of  activation of the channel  \cite{cgabc}. Therefore, a HH model of  a Na+ channel
may be expressed as a kinetic scheme that is consistent with the ion channel structure and the energy
landscape of each S4 sensor during activation and  inactivation processes. Although  it is often
assumed that the independence of Na+ channel  inactivation and activation is required
for the Na+ channel conductance expression  $m^3 h$  \cite{hille}, strongly coupled 
activation and inactivation is also compatible with the open state probability 
$O(t) = m(t)^3 h(t)$.

 {\bf  REDUCTION OF A KINETIC MODEL FOR Na+ CHANNEL ACTIVATION,  AND FAST AND SLOW   INACTIVATION}

In this section, it is assumed that  the activation of three voltage sensors  regulating the  Na+ channel 
 conductance is coupled to a two-stage inactivation process, and that slow inactivation is accessible from 
 fast inactivated states \cite{osijb}, and therefore, the kinetics may be described by a 
fifteen state model (see Fig. 11)
\begin{eqnarray}
\frac{dC_1}{dt} & = &  -(\alpha_{i1} + \alpha_{C1})C_1(t)  + \beta_{C1} C_2(t) + \beta_{i1} A_1(t) \label{15c1} \\
\frac{dC_2}{dt} & = &  -(\alpha_{i2} + \alpha_{C2} + \beta_{C1})C_2(t)  + \alpha_{C1} C_1(t) + \nonumber  \\
                &   &  \beta_{C2}C_3(t) + \beta_{i2} A_2(t) \label{15c2} \\
\frac{dC_3}{dt} & = &  -(\alpha_{i3} + \alpha_O + \beta_{C2})C_3(t)  + \alpha_{C2} C_2(t) + \nonumber  \\
                &   &  \beta_O O(t) + \beta_{i3} A_3(t) \label{15c3} \\
\frac{dO}{dt} & = &   - (\beta_O + \alpha_{i4})O(t) + \alpha_O C_3(t) + \beta_{i4} A_4(t) \label{15o} \\
\frac{dA_1}{dt} & = &   -(\alpha_{A1} + \beta_{i1} + \gamma_{i1})A_1(t) + \alpha_{i1} C_1(t) \nonumber \\
               &   & + \delta_{i1} I_1(t) + \beta_{A1} A_2(t) \label{15a1} \\
\frac{dA_2}{dt} & = &  -(\alpha_{A2} + \beta_{A1} + \beta_{i2} + \gamma_{i2})A_2(t) + \alpha_{i2} C_2(t) \nonumber \\
                &   & +\delta_{i2}I_2(t) + \alpha_{A1} A_1(t) + \beta_{A2} A_3(t) \label{15a2} \\
\frac{dA_3}{dt} & = &  -(\alpha_{A3} + \beta_{A2} + \beta_{i3} + \gamma_{i3})A_3(t) + \alpha_{i3} C_3(t) \nonumber \\
                &   & +\delta_{i3}I_3(t) + \alpha_{A2} A_2(t) + \beta_{A3} A_4(t) \label{15a3} \\
\frac{dA_4}{dt} & = &  -(\beta_{A3} + \beta_{i4} + \gamma_{i4})A_4(t) + \alpha_{i4} O(t) \nonumber \\
                &   & +\delta_{i4}I_4(t) + \alpha_{A3} A_3(t)  \label{15a4} \\
\frac{dI_1}{dt} & = & -(\alpha_{I1} + \delta_{i1}) I_1(t) + \gamma_{i1} A_1(t) +\beta_{I1} I_2(t) \label{15i1} \\
\frac{dI_2}{dt} & = &  -(\alpha_{I2} + \beta_{I1} + \delta_{i2} + \mu) I_2(t) \nonumber  \\
                &   &  + \gamma_{i2} A_2(t) +\alpha_{I1} I_1(t) +\beta_{I2} I_3(t) + \nu S_2(t), \label{15i2} \\
\frac{dI_3}{dt} & = &   -(\alpha_{I3} + \beta_{I2} + \delta_{i3}+ \mu) I_3(t) \nonumber  \\
		&   &  + \gamma_{i3} A_3(t) +\alpha_{I2} I_2(t)+ \beta_{I3} I_4(t) + \nu S_3(t) \label{15i3}  \\
\frac{dI_4}{dt} & = &  - ( \beta_{I3} + \delta_{i4}+ \mu) I_4(t)     \nonumber  \\ 
		&   &		 +\gamma_{i4} A_4(t) + \alpha_{I3} I_3(t) +  \nu S_4(t)   \label{15i4} \\
\frac{dS_2}{dt} & = & -(\alpha_{I2} + \nu) S_2(t)  +\beta_{I2} S_3(t) + \mu I_2(t) \label{15s2}  \\
\frac{dS_3}{dt} & = & -(\alpha_{I3} +\beta_{I2} + \nu) S_3(t)  + \alpha_{I2} S_2(t) + \beta_{I3} S_4(t) \nonumber  \\
		&   &  + \mu I_3(t) \label{15s3} \\
\frac{dS_4}{dt} & = & -(\beta_{I3} + \nu) S_4(t)  + \alpha_{I3} S_3(t) + \mu I_4(t), \label{15s4}
\end{eqnarray}
where $S_2(t)$, $S_3(t)$ and $S_4(t)$ are the occupational probabilities for the slow inactivated states, 
and  $\mu$ and $\nu$ are voltage dependent transition rates.
 As $I_1(t) \approx 0$ following a transient,
it may be assumed  that entry into the slow inactivated state corresponding to $I_1$ is also small,
and has no effect on the dynamics. By  expressing the solution as a two-scale asymptotic expansion and
eliminating secular terms \cite{kc}, Eqs. (\ref{15c1}) to (\ref{15s4}) may be reduced to
an eleven state system when the two-stage inactivation transitions satisfy 
$ \alpha_{ik} \ll \gamma_{ik}$, $\delta_{ik} \ll \beta_{ik}$ and   $\gamma_{ik} + \beta_{ik}$
is greater than the activation and deactivation rate functions, for each $k$ \cite{vac1} 
(see Fig. 12)  
\begin{eqnarray}
\frac{dC_1}{dt} & = &  -(\rho_1 + \alpha_{C1})C_1(t) + \beta_{C1} C_2(t) + \sigma_1 I_1(t) \label{8s1} \\
\frac{dC_2}{dt} & = &   - (\alpha_{C2} + \beta_{C1} + \rho_2) C_2(t) + \alpha_{C1} C_1(t) + \beta_{C2} C_3(t) + \sigma_2 I_2(t) \label{8s2} \\
\frac{dC_3}{dt} & = &   - (\alpha_{O} + \beta_{C2} + \rho_3) C_3(t) + \alpha_{C2} C_2(t) + \beta_{O} O(t) + \sigma_3 I_3(t) \label{8s3} \\
\frac{dO}{dt}   & = &    - (\beta_O + \rho_4) O(t) + \alpha_O C_3(t) + \sigma_4 I_4(t)  \label{8s4} \\
\frac{dI_1}{dt} & = &   -(\alpha_{I1} + \sigma_1) I_1(t) + \rho_1 C_1(t) + \beta_{I1} I_2(t)  \label{8s5} \\
\frac{dI_2}{dt} & = &    -(\alpha_{I2} +\beta_{I1} + \sigma_2 + \mu) I_2(t) + \alpha_{I1}I_1(t) + \beta_{I2} I_3(t)  + \nonumber  \\
		&   & \rho_2 C_2(t) + \nu S_2(t)  \label{8s6} \\	
\frac{dI_3}{dt} & = &     -(\alpha_{I3} +\beta_{I2} + \sigma_3 + \mu) I_3(t) + \alpha_{I2}I_2(t) + \beta_{I3} I_4(t) + \nonumber  \\
                &   &    \rho_3 C_3(t) + \nu S_3(t),  \label{8s7} \\
\frac{dI_4}{dt} & = &     - (\beta_{I3} + \sigma_4 + \mu) I_4(t)  + \alpha_{I3}  I_3(t) + \rho_4 O(t) + \nu S_4(t),  \label{8s8} \\
\frac{dS_2}{dt} & = &    -(\alpha_{I2} + \nu) S_2(t)  + \beta_{I2} S_3(t) + \mu I_2(t) \label{8s9} \\
\frac{dS_3}{dt} & = &    -(\alpha_{I3} + \beta_{I2} + \nu) S_3(t)  + \alpha_{I2}  S_2(t) + \beta_{I3} S_4(t) + \mu I_3(t) \label{8s10} \\
\frac{dS_4}{dt} & = &    -(\beta_{I3} + \nu) S_4(t) + \alpha_{I3} S_3(t) + \mu I_4(t). \label{8s11}
\end{eqnarray}  
 Assuming that $\alpha_{I1} \gg \rho_1 $  and  $ \sigma_1 \gg \beta_{I1}$ \cite{cgabc}, 
Eqs. (\ref{8s1}) and (\ref{8s6}) may be expressed as 
\begin{eqnarray}
\frac{dC_1}{dt} & = &  -(\alpha_{C1} + \hat{\rho}_1) C_1(t) + \beta_{C1} C_2(t) + \hat{\sigma}_1 I_2(t)
 \label{8s1a} \\
\frac{dI_2}{dt} & = &  - (\alpha_{I2} + \hat{\sigma}_1 + \sigma_2 + \mu) I_2(t) + \beta_{I2} I_3(t) + \nu S_2(t) \nonumber  \\ 
		&   &   +\hat{\rho}_1  C_1(t) + \rho_2 C_2(t), \label{8s5a} 
\end{eqnarray}
where $\hat{\rho}_1$ and $\hat{\sigma}_1$ are defined in Eqs. (\ref{rho1}) and (\ref{sig1}), and
the kinetics may be represented by the ten state model in Fig. 13.

In Eqs. (\ref{8s7}) to (\ref{8s11}) and (\ref{8s5a}), it is assumed that the transition rates
between fast inactivated states $I_2$, $I_3$ and  $I_4$, and between slow inactivated states
$S_2$, $S_3$ and  $S_4$,  are an order of magnitude larger than the corresponding 
inactivation and recovery rates, and the activation  and deactivation rates between
closed and open states, and  therefore, by expressing the solution 
as a three-scale asymptotic  expansion and eliminating secular terms \cite{kc}, 
Eqs. (\ref{8s1})  to  (\ref{8s11})  are reducible to a six state kinetic model (see Fig. 14) 
\begin{eqnarray}
\frac{dC_1}{dt} & = &  -(\rho_1 + \alpha_{C1})C_1(t) + \beta_{C1} C_2(t) +\hat{\sigma}_{1r}  I(t)     \label{ss1} \\
\frac{dC_2}{dt} & = &   - (\alpha_{C2} + \beta_{C1} + \rho_2) C_2(t) + \alpha_{C1} C_1(t) + \beta_{C2} C_3(t) + \sigma_{2r}  I(t) \label{ss2} \\
\frac{dC_3}{dt} & = &  - (\alpha_{O} + \beta_{C2} + \rho_3) C_3(t) +  \alpha_{C2} C_2(t) + \beta_{O} O(t)  + \sigma_{3r}  I(t)  \label{ss3} \\
\frac{dO}{dt}    & = &    - (\beta_O + \rho_4) O(t) +  \alpha_O C_3(t) + \sigma_{4r} I(t)  \label{ss4} \\
\frac{dI}{dt}    & = &   - (\hat{\sigma}_{1r} + \sigma_{2r} +  \sigma_{3r} +  \sigma_{4r} + \mu) I(t)  \nonumber  \\
                 &    &   +\hat{\rho}_1 C_1(t) + \rho_2 C_2(t) + \rho_3 C_3(t) + \rho_4 O(t) + \nu S(t) \label{ss5} \\
\frac{dS}{dt} & = &    \mu I(t) - \nu S(t) \label{ss6}
\end{eqnarray}
where $\hat{\sigma}_{1r}$, $\sigma_{2r}$, $\sigma_{3r}$ and $\sigma_{4r}$ are defined in 
Eqs. (\ref{sig1r})  to  (\ref{sig4r}),  $C_1(t) + C_2(t) + C_3(t)+O(t) + I(t) + S(t)= 1$, and
 following a transient, the inactivation probabilities $I_2(t)$,  $I_3(t)$ and $I_4(t)$   may be
 approximated by Eqs. (\ref{i2}) to (\ref{i4}), and the slow inactivation probabilities 
$S_2(t)$, $S_3(t)$ and $S_4(t)$ may be expressed as 
\begin{eqnarray}
S_2(t)    & \approx &  \frac{\beta_{I2} \beta_{I3} S(t) }{\alpha_{I2}\alpha_{I3} + \alpha_{I2}\beta_{I3} +\beta_{I2} \beta_{I3}}    \label{s2} \\
S_3(t)   & \approx & \frac{\alpha_{I2} \beta_{I3} S(t) }{\alpha_{I2}\alpha_{I3} + \alpha_{I2}\beta_{I3} +\beta_{I2} \beta_{I3}}  \label{s3} \\
S_4(t)   & \approx & \frac{\alpha_{I2} \alpha_{I3} S(t) }{\alpha_{I2}\alpha_{I3} + \alpha_{I2}\beta_{I3} +\beta_{I2} \beta_{I3}}  \label{s4},
\end{eqnarray}
where $S(t) = S_2(t) + S_3(t) + S_4(t)$.

  Expressing $C_1(t) = m_1(t) h(t)$, $C_2(t) = m_2(t) h(t)$,$C_3(t) = m_3(t) h(t)$, $O(t) = m_O(t) h(t)$ and
 $h(t) = 1 - I(t) - S(t)$, where $m_1(t)$, $m_2(t)$, $m_3(t)$ and $m_O(t)$ are activation variables and $h(t)$
 is an inactivation variable, and assuming that  the  activation sensors are independent 
($\alpha_{C1} = 3 \alpha_m, \alpha_{C2} = 2 \alpha_m, \alpha_O =  \alpha_m, \beta_{C1} =  \beta_m,
 \beta_{C2} = 2 \beta_m, \beta_{O} = 3 \beta_m $), and that 
the inactivation rates are an order of magnitude smaller than the activation rates in  Eqs. (\ref{ss1})  to 
 (\ref{ss4}), it may be shown that $m_1(t) = (1 - m(t))^3$, $m_2(t) =  3 m(t)(1 - m(t))^2$, 
$m_3(t) =  3 m(t)^2(1 - m(t))$, $m_O(t) = m(t)^3$,  where $m(t)$ satisfies Eq. (\ref{mm3}),  $h(t)$ and
$S(t)$ satisfy
\begin{equation}
\frac{dh}{dt}        =   \alpha_h ( 1 - S(t)) - h(t)(\alpha_h + \beta_h)   \label{rateh} 
\end{equation}
\begin{equation}
\frac{dS}{dt}        =  \mu ( 1 - h(t)) - S(t)(\mu  + \nu),  \label{rates} 
\end{equation}
and $\beta_h$ and $\alpha_h$ are defined in Eqs. (\ref{beth4})  and  (\ref{alfh4}).

Defining total inactivation $T(t) = I(t) + S(t) = 1 - h(t)$, Eqs. (\ref{rateh})  and  (\ref{rates}) may be expressed as
\begin{equation}
\frac{dT}{dt}        =   \beta_h + \alpha_h S(t) - T(t)(\alpha_h + \beta_h)   \label{ratet} 
\end{equation}
\begin{equation}
\frac{dS}{dt}        =  \mu  T(t) - S(t)(\mu  + \nu).   \label{ratesi} 
\end{equation}
It is assumed that the K+  and leakage channels repolarize the membrane, and if the K+ conductance is
proportional to $n(t)^j$ where $j$ is the number of voltage sensors such that $1 \le j \le 4$,
and the activation variable $n(t)$ satisfies Eq. (\ref{nn}), the membrane current equation is
\begin{equation}
C\frac{dV}{dt}= i_e - \bar{g}_{Na} O(t) (V - V_{Na}) - \bar{g}_{K}n(t)^j (V-V_{K}) - \bar{g}_{L} (V - V_{L}),  \label{cur4s}
\end{equation}
where $O(t) = m(t)^3 (1 - T(t))$.

The variable  $S(t)$ has a slow variation relative to fast inactivation, and therefore, writing 
$h(t)=h_{f}(t)(1-S(t))$, Eqs. (\ref{rateh}) and (\ref{rates}) may be expressed as   
\begin{equation}
\frac{dh_{f}}{dt}        \approx   \alpha_h  - h_{f}(t)(\alpha_h + \beta_h)   \label{ratehf} 
\end{equation}
\begin{equation}
\frac{dS}{dt}        =  \mu ( 1 - h_f(t)) - S(t)[\mu ( 1 - h_f(t))  + \nu],  \label{ratesf} 
\end{equation}
and the forward rate for slow inactivation is dependent on the fast inactivation variable $h_{f}(t)$, similar 
to the dependence of the fast inactivation rate $\rho(t) \approx \beta_h$ on the activation variable $m(t)$
in Eq. (\ref{rhosum3}). Defining $s(t)=1-S(t)$,   Eq. (\ref{ratesf})  is equivalent to
\begin{equation}
\frac{ds}{dt}        =  \nu - s(t)[\nu + \mu( 1 - h_f(t)) ].  \label{ratesf2} 
\end{equation}

During a voltage clamp of the Na+ channel membrane,  $h_{f}(t)$ may be 
approximated by $h_{f\infty}(V) = \alpha_h/(\alpha_h + \beta_h)$ following the relaxation of the 
fast inactivation process,  and from Eq. (\ref{ratesf2}), we may write
\begin{equation}
\frac{ds}{dt}        =  \alpha_s - s(t)(\alpha_s + \beta_s),  \label{slow3} 
\end{equation}
where $\alpha_s = \nu$ and $ \beta_s \approx \mu( 1 - h_{f \infty})$. If $\mu$ has a weak
voltage dependence, there is a plateau in the voltage dependence of $\beta_s$ for a large depolarization
potential, consistent with the slow inactivation voltage clamp data for a Na+ channel \cite{ffg}.
Therefore, Eq. (\ref{cur4s}) may be expressed as
\begin{equation}
C\frac{dV}{dt}= i_e - \bar{g}_{Na} m^3 h_{f} s (V - V_{Na}) - \bar{g}_{K}n^j(V-V_{K}) - \bar{g}_{L} (V - V_{L}),
  \label{cur4s2}
\end{equation}
and Eqs.  (\ref{nn}), (\ref{mm3}), (\ref{ratehf}),  (\ref{slow3}),  and (\ref{cur4s2}) are the empirical 
equations that describe spike frequency adaptation \cite{ffg}.

The variation in the probability $S$ that the inactivation sensor occupies a slow inactivation state is an order
of magnitude slower than for the fast inactivation probability I, and S may be treated as a parameter
that modifies the stability of the stationary state of the $(m,n,T,V)$  subsystem. 
During a spike train, the increase in the value of the slow inactivation variable S is associated with
a delay to the next spike,  and when the stationary state of the  subsystem becomes stable,  the system 
returns to the resting potential. The solution of  Eqs. (\ref{15c1})  to  (\ref{15s4}) may be approximated
 by the solution of Eqs. (\ref{mm3}),   (\ref{ratet}) and (\ref{ratesi}) where n and V are determined by 
 Eqs. (\ref{nn}) and  Eq. (\ref{cur4s2}) (see Fig. 15).

 A similar process occurs during a repetitive bursting oscillation where slow inactivation increases until the
 stationary  state of the subsystem becomes stable; however, in this case, as the slow variable relaxes during 
the  subthreshold oscillation, the stationary state of the subsystem loses its stability  when the
recovery rate $\nu$ for slow  inactivation is sufficiently large, and the bursting oscillation 
resumes \cite{wc} (see Fig. 16). More generally, bursting activity serves an important role in the
 nervous system and in addition to the slow inactivation of Na+ channels, may also be generated by
inactivating K+ channels and the activation of slow M-type K+ channels \cite{izhik}.
Although the cardiac ventricular action potential is dependent on 
Na+, K+ and Ca++ currents as well as intracellular ion concentration changes \cite{cr}, for a simplified model
of the action potential that is dependent only on Na+, K+ and leakage currents,  if the rate of recovery
 from Na+ channel fast inactivation is increased, the stationary state  of the subsystem  is stable for small 
values of  $S$, but may lose its stability as S increases and therefore, the plateau may develop an oscillation 
 (see Figs. 17 and 18). 

\newpage
   {\bf CONCLUSION}

Based on an empirical description of the voltage clamp K+ and Na+ channel currents and the calculation
of the membrane potential  from the ion current equation, the HH model accounts for subthreshold 
oscillations and the  action potential in the squid axon membrane \cite{hh}. 
The slow cumulative adaptation of spike firing during prolonged depolarization is associated with
a reduction in the number of Na+ channels available for activation,  and the Na+ current may be described 
by the expression $m^3 hs(V_{Na} - V)$ where the HH  equations for Na+ activation $m$ and fast
inactivation $h$ are supplemented by an independent rate equation for the slow inactivation
variable $s$ \cite{ffg}. However, recently it has been proposed that fast and slow
Na+ channel inactivation are sequential processes, and therefore, fast and slow inactivation are mutually 
dependent  \cite{osijb}.

In this paper, it has been shown that during an action potential, for a Na+ channel with three activation
sensors coupled to a two-stage inactivation process,  by expressing the solution as a two-scale asymptotic
 expansion and eliminating secular terms, a twelve state kinetic model  may be reduced to a seven state 
system when the first forward and backward inactivation transitions are rate limiting, and the recovery
 rate for fast inactivation $\sigma_1$ from the first inactivated state $I_1$ is an order of magnitude larger 
than the deactivation rate $\beta_{I1}$. If the transition rates between the fast inactivated
 states $I_2$ and $I_4$ are larger then the corresponding inactivation and recovery rates, and the occupation 
probabilities for closed  states and the  open state are expressed in terms of activation and fast inactivation
variables,  the model may  be further reduced to a system of  equations in the activation variables
 $m_1$, $m_2$, $m_3$, $m_O$ and the inactivation variable $h$. Assuming that the activation sensors are 
mutually independent, the twelve state kinetic model may be reduced to equations for m and h, and
therefore,  decreases the computation time for the simulation of action 
potentials.

The rate of recovery from inactivation $\alpha_h$ is dependent on 
the rate functions $\alpha_{I1}$ and $\beta_{I1}$, and the recovery rate
 $\sigma_1$, as $\sigma_{2}, \sigma_{3}, \sigma_{4} \ll  \sigma_{1}$,
 but for a moderate  hyperpolarization, the voltage dependence of $\alpha_h$ may be approximated 
by the exponential function $\beta_{I1}$,  in agreement with experimental studies on 
Na+ channel gating \cite{hh,cgabc,kb}. Assuming that the activation sensors are mutually independent,
the expression for the inactivation rate $\rho(t)$ is dependent on  $m(t)$, and the forward transition rates 
$\rho_k$ of the DIV sensor, and  if $m(t)$ has a faster relaxation than $h(t)$, $\rho(t)$ may 
be approximated by a voltage dependent function $\beta_h$, as assumed by HH \cite{hh}.
Therefore, the parameters for inactivation  may be calculated from the transition rates of the 
kinetic model based on the  structure of the wild-type or mutant Na+ ion channel. 
However,  the inactivation rates $\rho_k$ and recovery rates  $\sigma_k$  are an order of magnitude 
smaller than activation and deactivation rates, and it may be shown that $\rho_k$, $\sigma_k$
and  the inactivation variable  $h(t)$ generally  only have a small effect on the time-dependence
of the activation variable $m(t)$. 

If  the Na+ channel permits a slow transition to  additional inactivated states, by expressing the solution as
a three-scale asymptotic  expansion and eliminating secular terms,  the kinetic model for channel gating
may be reduced to a six state system of equations when, in addition to the conditions
satisfied by the fast inactivation rates,  the transition rates between  the slow inactivated states
are an order of magnitude  larger than the slow inactivation and recovery rates. Assuming that
the activation sensors are mutually independent, a fifteen state kinetic model of  Na+ channel gating  
may be reduced to  equations for  activation,  and fast and slow inactivation that approximate
the empirical equations  that describe spike frequency adaptation in a neural membrane, 
a repetitive bursting oscillation that is modulated by the slow inactivation of Na+ channels,
and a plateau oscillation during a cardiac action potential.

\newpage

\begin{figure*}
\begin{center}
\includegraphics[width=0.8 \textwidth]{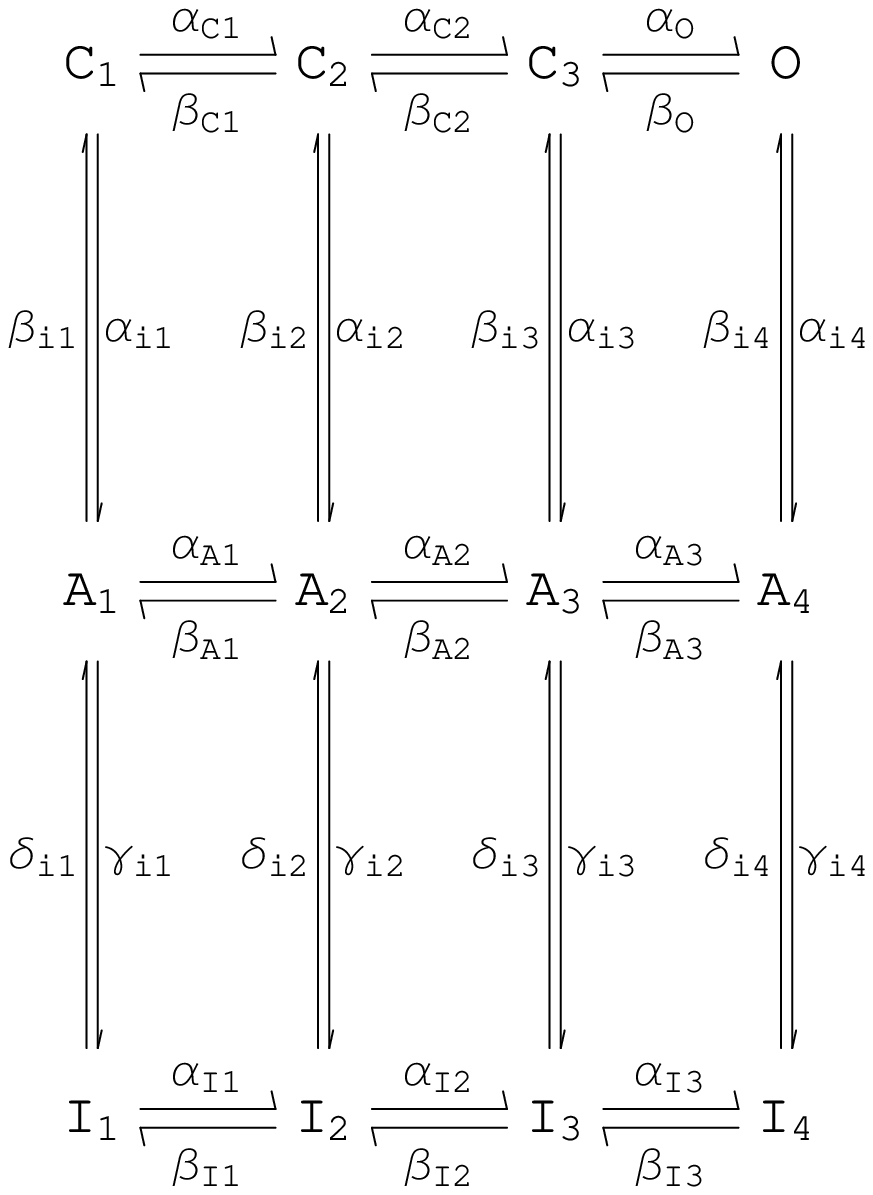}
\caption{
State diagram for Na+ channel gating where horizontal transitions represent the activation of three voltage sensors
 (DI, DII and DIII)  that open the pore, and vertical transitions represent the two stage fast inactivation 
process of the DIV voltage sensor and the inactivation motif.
}
\end{center}
\end{figure*}

\begin{figure*}
\begin{center}
\includegraphics[width=0.8 \textwidth]{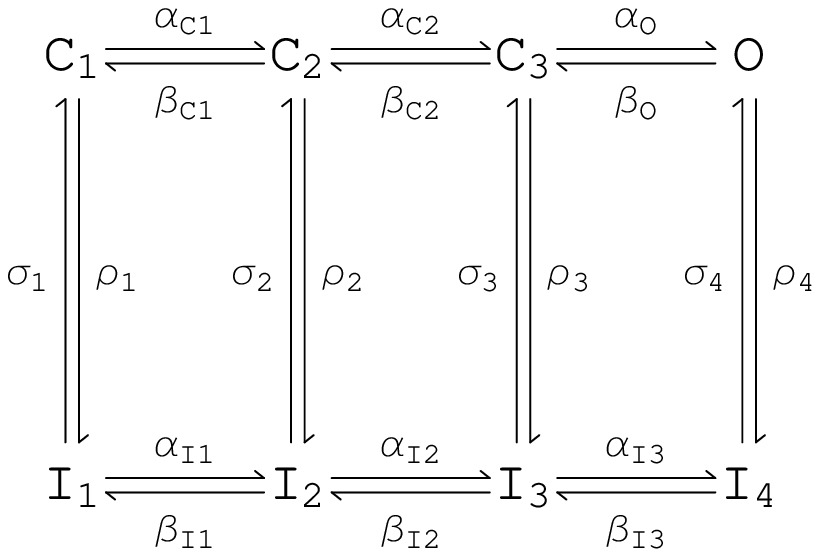}
\caption{
The state diagram for Na+ channel gating in Fig. 1 may be reduced to an eight-state model when 
$\beta_{ik} \gg \delta_{ik}$, $\gamma_{ik} \gg \alpha_{ik}$, 
and  $\gamma_{ik} + \beta_{ik}$  is greater than the activation and deactivation rate functions,
for each $k$,  where the derived rate functions $\rho_k$ and $\sigma_k$ are defined in 
Eqs. (\ref{rhok}) and (\ref{sigk}).
}
\end{center}
\end{figure*}

\begin{figure*}
\begin{center}
\includegraphics[width=0.8 \textwidth]{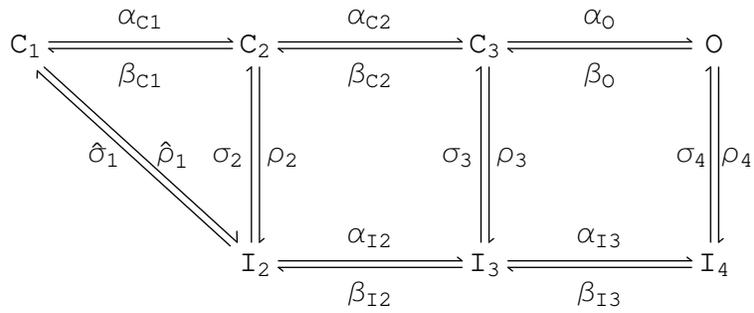}
\caption{
The state diagram for Na+ channel gating in Fig. 2 may be reduced to a seven-state model when 
$\alpha_{I1} \gg \rho_1$ and $\sigma_1 \gg \beta_{I1}$,  where the  derived rate functions
$\hat{\rho}_1$ and $\hat{\sigma}_1$ are defined in Eqs. (\ref{rho1}) and (\ref{sig1}).
}
\end{center}
\end{figure*}

\begin{figure*}
\begin{center}
\includegraphics[width=0.8 \textwidth]{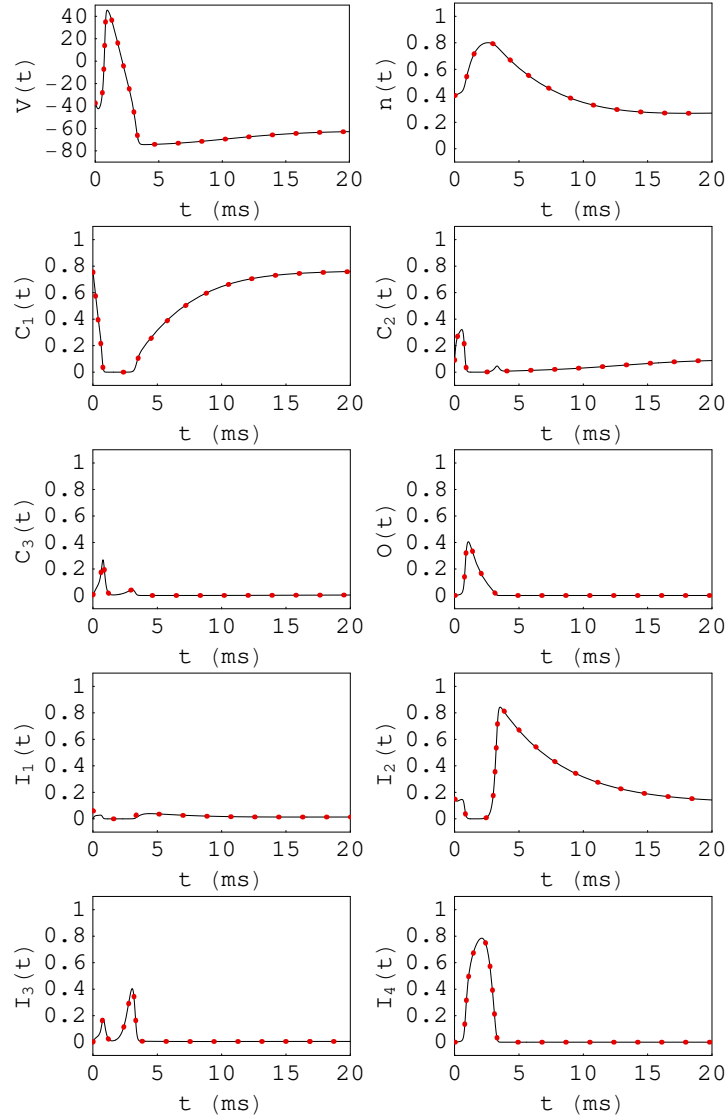}
\caption{
For the solution of  Eqs. (\ref{8c1}) to (\ref{8c8}) during an action potential  (solid line), Eqs. 
(\ref{8c1}), (\ref{8c5}) and (\ref{8c6}) may be approximated  by  Eqs. (\ref{8c1a}),  (\ref{8c6a}) and
 (\ref{8c5a}), (dotted line), when $\alpha_{I1} \gg \rho_1$ and $\sigma_1 \gg \beta_{I1}$, 
and n and V are determined by  Eqs. (\ref{nn}) and (\ref{cur3}). 
The rate functions are $\alpha_{m} = 0.1(V+35)/(1 - \exp[-(V+35)/10])$,
$\beta_m = 4 \exp[-(V+60)/18]$,  $\alpha_{C1} = 3 \alpha_{m }$, $\beta_{C1} = \beta_m$, 
$\alpha_{C2} = 2 \alpha_{m }$, $\beta_{C2} = 2 \beta_m$, $\alpha_{O} = \alpha_{m }$, 
$\beta_O = 3 \beta_m$, $ \alpha_{I1} =  \alpha_{C1}$, $\beta_{I1} = 0.016 \beta_{C1}$,
 $ \alpha_{I2} =  \alpha_{C2}$, $\beta_{I2} =  \beta_{C2}$, $ \alpha_{I3} = \alpha_{O}$,
$\beta_{I3} = \beta_O$, $\alpha_{ik} = 1$, $\gamma_{ik} = 22.2$, $\beta_{ik} = \exp[-V/10]$, 
$\delta_{i1} = 2.5$,  $\delta_{i2} = \delta_{i3} = \delta_{i4} = 0.04$, 
$\rho_k =  \alpha_{ik}/(1 + \beta_{ik}/\gamma_{ik})$ for $k = 1,4$, 
$\sigma_1 = \delta_{i1}/(1 +  \gamma_{i1}/\beta_{i1})$, 
$\sigma_2 = \sigma_3 = \sigma_4  = 0.016 \sigma_1$ (ms$^{-1}$)
$\alpha_{n} = 0.01(V+50)/(1 - \exp[-(V+50)/10])$, $\beta_n = 0.125 \exp[-(V+60)/80]$, and
$\bar{g}_{Na}$= 120 mS/cm$^2$, $\bar{g}_{K}$= 36 mS/cm$^2$, $\bar{g}_{L}$= 0.3 mS/cm$^2$,  $V_{Na}$ = 55 mV,
$V_{K}$ = -75 mV, $V_{L}$ = -60 mV, $C = 1$ $\mu$F/cm$^2$, $i_{e}$ = 1 $\mu$A/cm$^2$.
}
\end{center}
\end{figure*}

\begin{figure*}
\begin{center}
\includegraphics[width=0.8 \textwidth]{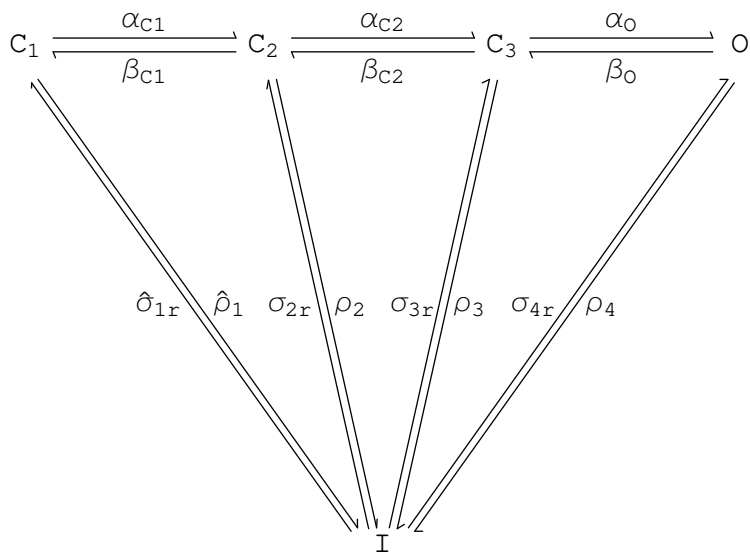}
\caption{
State diagram for Na+ channel gating in Fig. 3 may be reduced to a five state model when 
the transition rates between fast inactivated states are larger than inactivation and recovery rates.
}
\end{center}
\end{figure*}

\begin{figure*}
\begin{center}
\includegraphics[width=0.8 \textwidth]{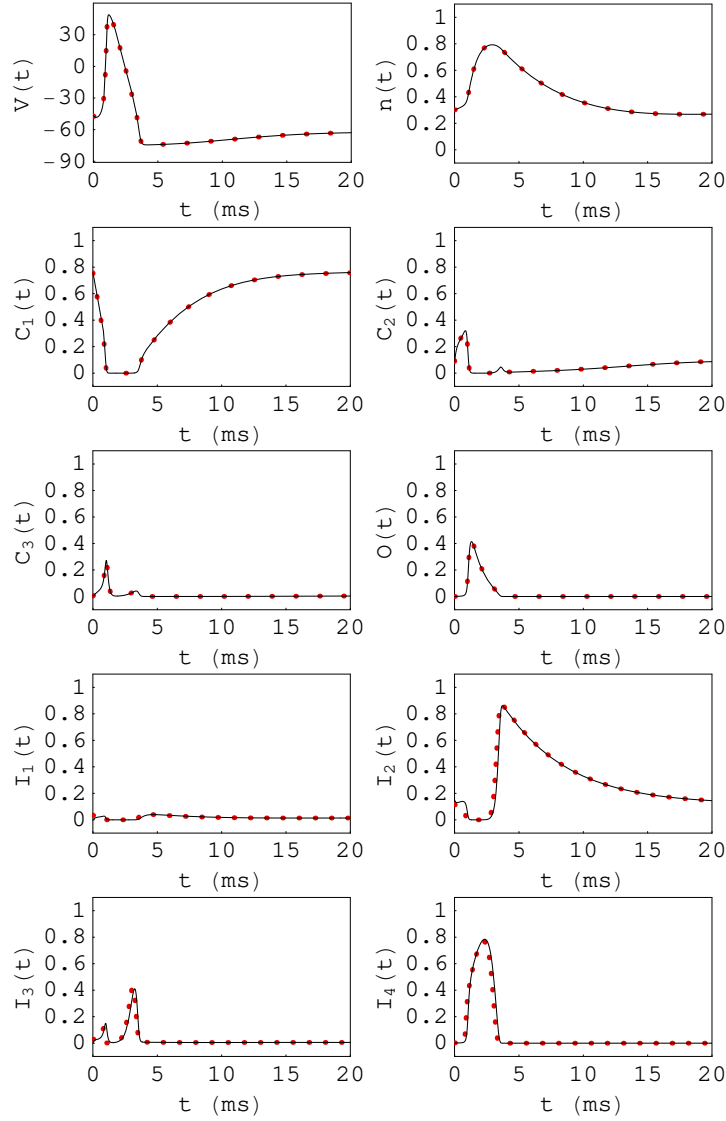}
\caption{
The solution of a  Na+ channel eight state kinetic model, Eqs. (\ref{8c1}) to (\ref{8c8})  (solid line)
may be approximated by the solution of Eqs. (\ref{is1}) to  (\ref{is5}) (dotted line), where n and V are determined by
Eqs.  (\ref{nn})  and  (\ref{cur3}), and $I_1$ to $I_4$ are calculated from Eqs. (\ref{8c5a}) and 
(\ref{i2}) to (\ref{i4}). 
The rate functions are $\alpha_{m} = 0.1(V+35)/(1 - \exp[-(V+35)/10])$,
$\beta_m = 4 \exp[-(V+60)/18]$,  $\alpha_{C1} = 3 \alpha_{m }$, $\beta_{C1} = \beta_m$, $\alpha_{C2} = 2 \alpha_{m }$,
$\beta_{C2} = 2 \beta_m$, $\alpha_{O} = \alpha_{m }$, $\beta_O = 3 \beta_m$, $ \alpha_{I1} =  \alpha_{C1}$, 
$\beta_{I1} = 0.016 \beta_{C1}$, $ \alpha_{I2} = 2 \alpha_{C2}$, $\beta_{I2} = 2 \beta_{C2}$, $ \alpha_{I3} = 2 \alpha_{O}$,
$\beta_{I3} = 2 \beta_O$, $\alpha_{ik} = 1$, $\gamma_{ik} = 22.2$, $\beta_{ik} = \exp[-V/10]$, 
$\delta_{i1} = 2.5$,  $\delta_{i2} = \delta_{i3} = \delta_{i4} = 0.04$, 
$\rho_k =  \alpha_{ik}/(1 + \beta_{ik}/\gamma_{ik})$ for $k = 1,4$, 
$\sigma_1 = \delta_{i1}/(1 +  \gamma_{i1}/\beta_{i1})$, 
$\sigma_2 = \sigma_3 = \sigma_4  = 0.016 \sigma_1$, $\alpha_{n} = 0.01(V+50)/(1 - \exp[-(V+50)/10])$,
$\beta_n = 0.125 \exp[-(V+60)/80]$ (ms$^{-1}$), and
$\bar{g}_{Na}$= 120 mS/cm$^2$, $\bar{g}_{K}$= 36 mS/cm$^2$, $\bar{g}_{L}$= 0.3 mS/cm$^2$,  $V_{Na}$ = 55 mV,
$V_{K}$ = -75 mV, $V_{L}$ = -60 mV, $C = 1$ $\mu$F/cm$^2$, $i_{e}$ = 1 $\mu$A/cm$^2$.
}
\end{center}
\end{figure*}

\begin{figure*}
\begin{center}
\includegraphics[width=0.4 \textwidth]{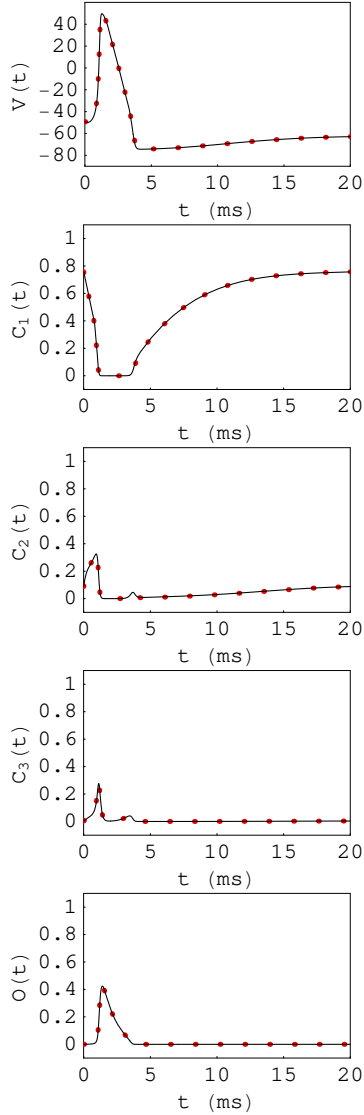}
\caption{
During an action potential, the solution of the Na+ channel activation equations, Eqs. (\ref{sm1}) to
(\ref{sm4})  (solid line) may be approximated by the solution of  Eqs. (\ref{sm1a}) to  (\ref{sm4a})
(dotted line), when the inactivation and recovery rates are an order of magnitude  smaller than 
the activation and deactivation rates, and n, V and h are determined by
 Eqs.  (\ref{nn}), (\ref{cur3}) and (\ref{sm5}). 
The rate functions are $\alpha_{m} = 0.1(V+35)/(1 - \exp[-(V+35)/10])$, $\beta_m = 4 \exp[-(V+60)/18]$,
  $\alpha_{C1} = 3 \alpha_{m }$, $\beta_{C1} = \beta_m$, $\alpha_{C2} = 2 \alpha_{m }$,
$\beta_{C2} = 2 \beta_m$, $\alpha_{O} =  \alpha_{m }$, $\beta_O = 3 \beta_m$, $ \alpha_{I1} = \alpha_{C1}$, 
$\beta_{I1} = 0.016 \beta_{C1}$, $ \alpha_{I2} = 2\alpha_{C2}$, $\beta_{I2} =  2\beta_{C2}$, $ \alpha_{I3} =  2\alpha_{O}$,
$\beta_{I3} =  2\beta_O$, $\alpha_{ik} = 1$, $\gamma_{ik} = 22.2$, $\beta_{ik} = \exp[-V/10]$,
$\rho_k =  \alpha_{ik}\gamma_{ik}/(\beta_{ik} + \gamma_{ik})$ for $k = 1,4$,
$\delta_{i1} = 2.5$,  $\delta_{i2} = \delta_{i3} = \delta_{i4} = 0.04$, 
$\sigma_1 = \delta_{i1}/(1 +  \gamma_{i1}/\beta_{i1})$, 
$\sigma_2 = \sigma_3 = \sigma_4  = 0.016 \sigma_1$,  $\alpha_{n} = 0.01(V+50)/(1 - \exp[-(V+50)/10])$,
$\beta_n = 0.125 \exp[-(V+60)/80]$ (ms$^{-1}$), and
$\bar{g}_{Na}$= 120 mS/cm$^2$, $\bar{g}_{K}$= 36 mS/cm$^2$, $\bar{g}_{L}$= 0.3 mS/cm$^2$,  $V_{Na}$ = 55 mV,
 $V_{K}$ = -75 mV, $V_{L}$ = -60 mV, $C = 1$ $\mu$F/cm$^2$, $i_{e}$ = 1 $\mu$A/cm$^2$.
}
\end{center}
\end{figure*}

\begin{figure*}
\begin{center}
\includegraphics[width=0.9 \textwidth]{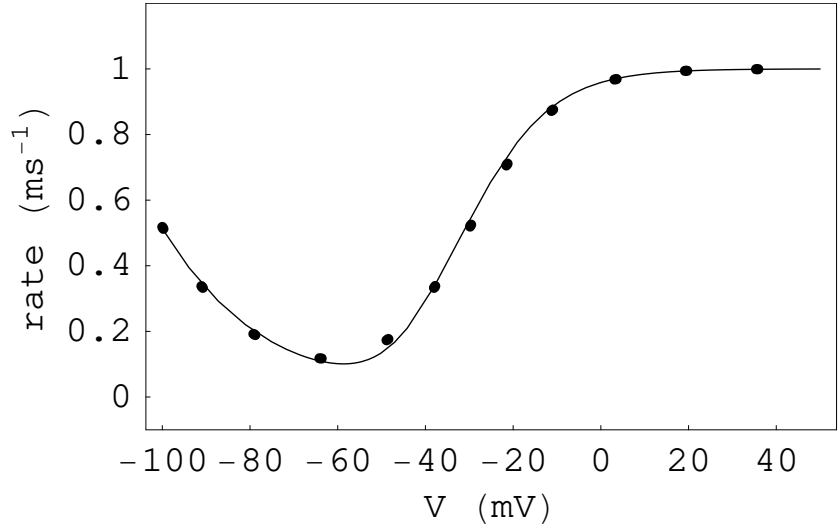}
\caption{
The voltage dependence of the Na+ channel HH inactivation rate function $\alpha_h + \beta_h$ (dotted line),
where $\alpha_h = 0.07 \exp[-(V + 60)/20]$  and $\beta_h = 1/(1+  \exp[-(V + 30)/10])$  may be
approximated by the expressions in Eqs.  (\ref{beth4}) and  (\ref{alfh4})   where the rate functions
are defined as $\alpha_{m} = 0.1(V+35)/(1 - \exp[-(V+35)/10])$, $\beta_m = 4 \exp[-(V+60)/18]$,
$\alpha_{C1} = 3 \alpha_{m }$, $\beta_{C1} = \beta_m$, $\alpha_{C2} = 2 \alpha_{m }$, 
$\beta_{C2} = 2 \beta_m$, $\alpha_{O} = \alpha_{m }$, $\beta_O = 3 \beta_m$, $ \alpha_{I1} = \alpha_{C1}$,
$\beta_{I1} = 0.016 \beta_{C1}$, $ \alpha_{I2} = 2\alpha_{C2}$, $ \beta_{I2} = 2\beta_{C2}$,
$ \alpha_{I3} = 2\alpha_{O}$,  $ \beta_{I3} = 2\beta_O$, $\alpha_{ik} = 1$, $\gamma_{ik} = 22.2$,
$\beta_{ik} = \exp[-V/10]$, $\delta_{i1} = 2.5$,  
$\delta_{i2} = \delta_{i3} = \delta_{i4} = 0.04$, $\rho_k =  \alpha_{ik}/(1 + \beta_{ik}/\gamma_{ik})$
for $k = 1,4$, $\sigma_1 = \delta_{i1}/(1 +  \gamma_{i1}/\beta_{i1})$, 
$\sigma_2 = \sigma_3 = \sigma_4  = 0.016 \sigma_1$ (ms$^{-1}$).
}
\end{center}
\end{figure*}

\begin{figure*}
\begin{center}
\includegraphics[width=0.8 \textwidth]{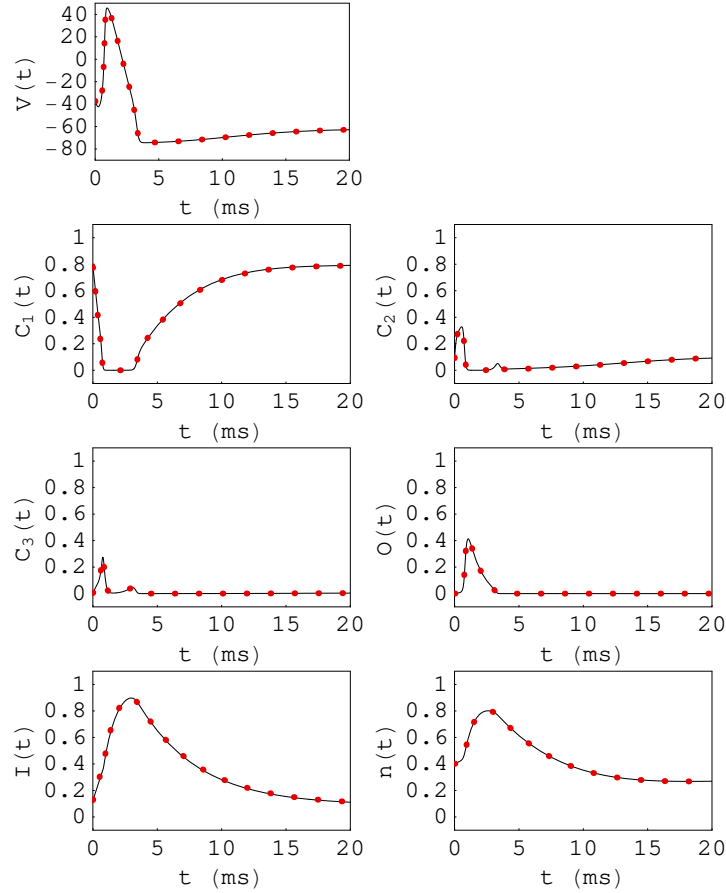}
\caption{ 
During an action potential, the solution of a  Na+ channel twelve state kinetic model,
 Eqs. (\ref{12c1}) to (\ref{12i4}) (solid line)  may be approximated by $C_1 = (1 - m)^3 h$,
$C_2 = 3m(1 - m)^2 h$, $C_3 = 3m^2(1 - m) h$, $O = m^3 h$, $I = 1 - h$ (dotted line), where 
$m$  and $h$ satisfy Eqs.  (\ref{mm3}) and (\ref{sm5a}), and $n$ and $V$  are determined by 
Eq. (\ref{nn}) and Eq. (\ref{cur3}). The conditions for the 
reduction are that (1) the two stage inactivation process satisfies $\beta_{ik} \gg \delta_{ik}$ and
 $\gamma_{ik} \gg \alpha_{ik}$, for each $k$ (see Fig. 1) (2) $\alpha_{I1} \gg \rho_1$ and $\sigma_1 \gg \beta_{I1}$
(see Fig. 2) and (3) the transition rates between inactivated states are an order of magnitude larger 
than inactivation and recovery rates (see Fig. 3). 
The rate functions are $\alpha_{m} = 0.1(V+35)/(1 - \exp[-(V+35)/10])$,
$\beta_m = 4 \exp[-(V+60)/18]$,  $\alpha_{C1} = 3 \alpha_{m }$, $\beta_{C1} = \beta_m$, 
$\alpha_{C2} = 2 \alpha_{m }$, $\beta_{C2} = 2 \beta_m$, $\alpha_{O} = \alpha_{m }$, 
$\beta_O = 3 \beta_m$, $ \alpha_{I1} =  \alpha_{C1}$,  $\beta_{I1} = 0.016 \beta_{C1}$,
 $ \alpha_{I2} = 2\alpha_{C2}$, $\beta_{I2} = 2\beta_{C2}$, $ \alpha_{I3} = 2\alpha_{O}$,
$\beta_{I3} = 2\beta_O$, $\alpha_{ik} = 1$, $\gamma_{ik} = 22.2$, $\beta_{ik} = \exp[-V/10]$,
 $\delta_{i1} = 2.5$,  $\delta_{i2} = \delta_{i3} = \delta_{i4} = 0.04$, 
$\rho_k =  \alpha_{ik}/(1 + \beta_{ik}/\gamma_{ik})$ for $k = 1,4$, 
$\sigma_1 = \delta_{i1}/(1 +  \gamma_{i1}/\beta_{i1})$, 
$\sigma_2 = \sigma_3 = \sigma_4  = 0.016 \sigma_1$, $\alpha_{n} = 0.01(V+50)/(1 - \exp[-(V+50)/10])$,
$\beta_n = 0.125 \exp[-(V+60)/80]$ (ms$^{-1}$), and $\bar{g}_{Na}$= 120 mS/cm$^2$, $\bar{g}_{K}$= 36 mS/cm$^2$,
$\bar{g}_{L}$= 0.3 mS/cm$^2$,  $V_{Na}$ = 55 mV, $V_{K}$ = -75 mV,
$V_{L}$ = -60 mV, $C = 1$ $\mu$F/cm$^2$, $i_{e}$ = 1 $\mu$A/cm$^2$.
}
\end{center}
\end{figure*}

\begin{figure*}
\begin{center}
\includegraphics[width=0.8 \textwidth]{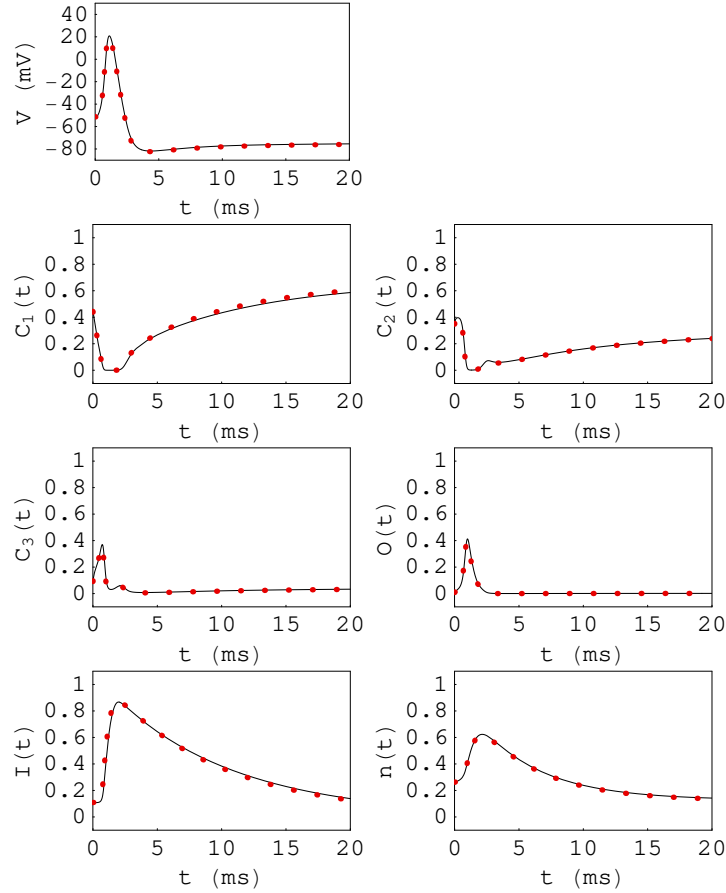}
\caption{ 
The solution of a  Na+ channel twelve state kinetic model, Eqs. (\ref{12c1}) to (\ref{12i4}) (solid line)
may be approximated by $C_1 = (1 - m)^3 h$,
$C_2 = 3m(1 - m)^2 h$, $C_3 = 3m^2(1 - m) h$, $O = m^3 h$, $I = 1 - h$ (dotted line), where 
$m$  and $h$ satisfy Eqs.  (\ref{mm3}) and (\ref{sm5a}), and $n$ and $V$  are determined by 
Eq. (\ref{nn}) and Eq. (\ref{cur3}).
The rate functions are $\alpha_{m} = 7.45 \exp[0.5V/25]$, $\beta_m = 0.8 \exp[-0.9V/25]$, 
$\alpha_{C1} = 3 \alpha_{m }$, $\beta_{C1} = \beta_m$, $\alpha_{C2} = 2 \alpha_{m }$,
$\beta_{C2} = 2 \beta_m$, $\alpha_{O} = \alpha_{m }$, $\beta_O = 3 \beta_m$, $ \alpha_{I1} = \alpha_{C1}$,
$\beta_{I1} = 0.01 \beta_{C1}$, $\alpha_{I2} =  2\alpha_{C2}$, $\beta_{I2} = 0.2 \beta_{C2}$, 
$\alpha_{I3} = 2\alpha_{O}$, $ \beta_{I3} = 0.2 \beta_O$,  
$\beta_{i1} = 2000 \exp[-2.4V/25]$, $\beta_{i2} = 200 \exp[-2.4V/25]$,  
$\beta_{i3} = 20 \exp[-2.4V/25]$, $\beta_{i4} = 2 \exp[-2.4V/25]$,
$\delta_{i1} = 1$,  $\delta_{i2} = \delta_{i3} = \delta_{i4} = 0.1$, 
$\alpha_{ik} = 2.1$, $\gamma_{ik} = 25$, 
$\rho_k =  \alpha_{ik}/(1 + \beta_{ik}/\gamma_{ik})$,  
$\sigma_k = \delta_{ik}/(1 +  \gamma_{ik}/\beta_{ik})$, 
for $k = 1,4$, $\alpha_{n} = 0.01(V+50)/(1 - \exp[-(V+50)/10])$, $\beta_n = 0.125 \exp[-(V+60)/80]$,
(ms$^{-1}$), and  $\bar{g}_{Na}$ = 20 mS/cm$^2$, $\bar{g}_{K}$ = 10 mS/cm$^2$, $\bar{g}_{L}$ = 1 mS/cm$^2$, 
$V_{Na}$ = 40 mV, $V_{K}$ = -90 mV, $V_{L}$  = -80 mV, 
$C = 1$ $\mu$F/cm$^2$, $i_{e}$ = 1 $\mu$A/cm$^2$.
}
\end{center}
\end{figure*}

\begin{figure*}
\begin{center}
\includegraphics[width=0.8 \textwidth]{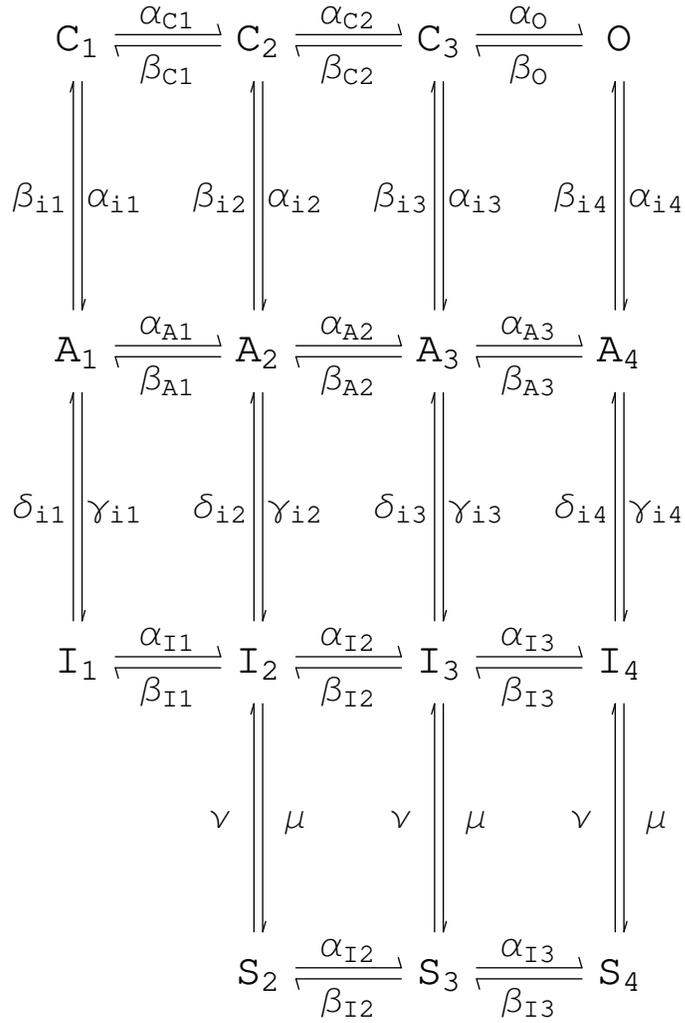}
\caption{
State diagram for Na+ channel gating where horizontal transitions represent the activation of
three voltage sensors  (DI, DII and DIII)  that open the pore, and vertical transitions represent
the two stage fast inactivation  process to states $I_1(t)$  to $I_4(t)$, and slow inactivation 
to states $S_2(t)$  to $S_4(t)$.
}
\end{center}
\end{figure*}

\begin{figure*}
\begin{center}
\includegraphics[width=0.6 \textwidth]{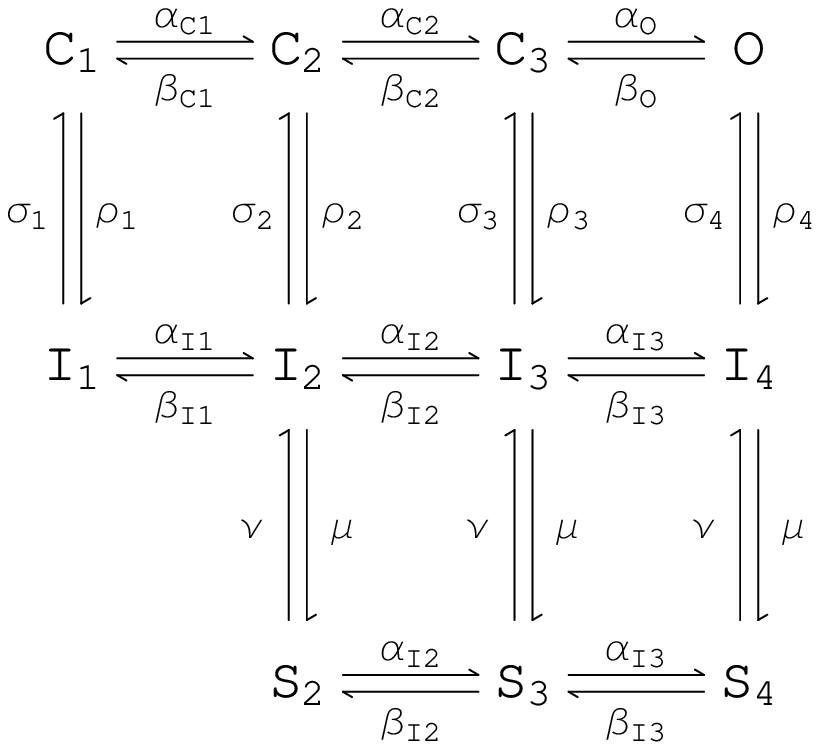}
\caption{State diagram for Na+ channel gating in Fig. 11 may be reduced to an eleven state model when 
$\beta_{ik} \gg \delta_{ik}$, $\gamma_{ik} \gg \alpha_{ik}$, and  $\gamma_{ik} + \beta_{ik}$  is greater
 than the activation and deactivation rate functions, for each $k$, where the
derived rate functions are $\rho_k$ and $\sigma_k$ defined in Eqs. (\ref{rhok}) and (\ref{sigk}).
}
\end{center}
\end{figure*}

\begin{figure*}
\begin{center}
\includegraphics[width=0.8 \textwidth]{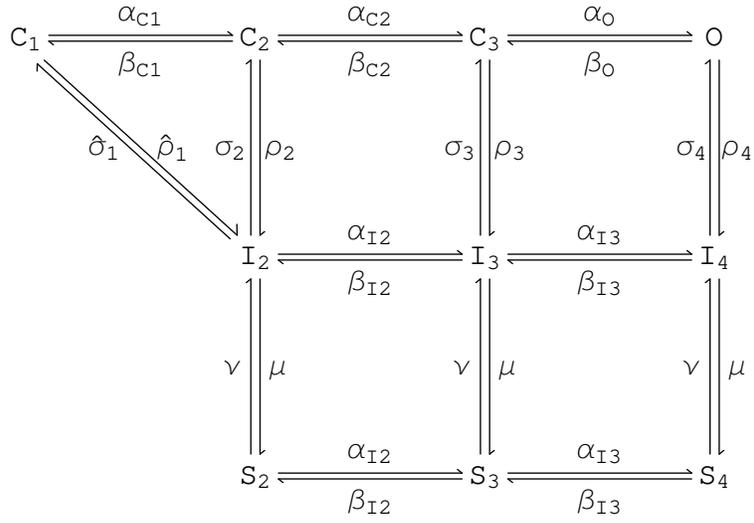}
\caption{The eleven state system for Na+ channel gating in Fig. 12 may be reduced to a 
ten state system when $\alpha_{I1} \gg \rho_1 $  and $ \sigma_1 \gg \beta_{I1}$, where the  derived
rate functions $\hat{\rho}_1$ and $\hat{\sigma}_1$ are defined in Eqs. (\ref{rho1}) and (\ref{sig1}).
}
\end{center}
\end{figure*}

\begin{figure*}
\begin{center}
\includegraphics[width=0.8 \textwidth]{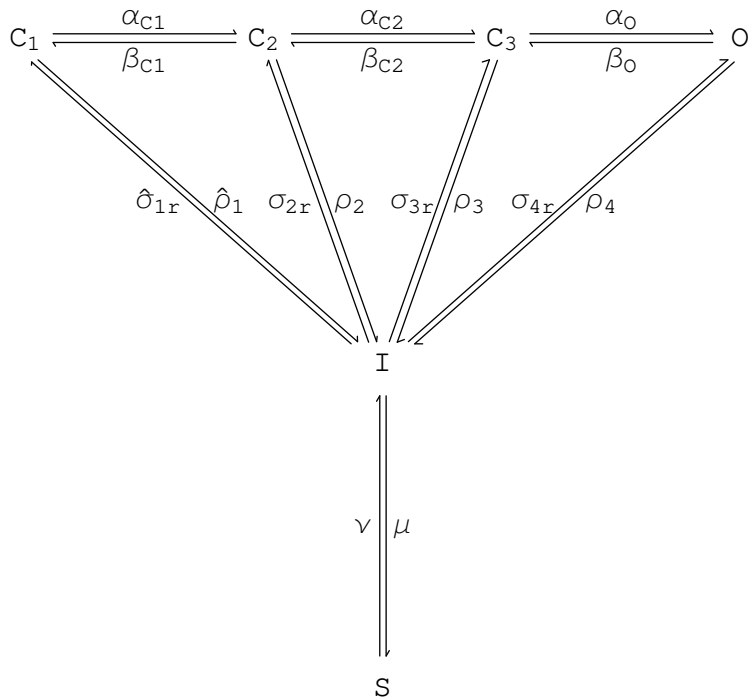}
\caption{The ten state system for Na+ channel gating in Fig. 13 may be reduced 
to a six state system when the transition rates between fast inactivated states $I_2(t)$ to $I_4(t)$,
 and between slow inactivated states $S_2(t)$ to $S_4(t)$ are larger than inactivation and recovery rates.
}
\end{center}
\end{figure*}

\begin{figure*}
\begin{center}
\includegraphics[width=0.7 \textwidth]{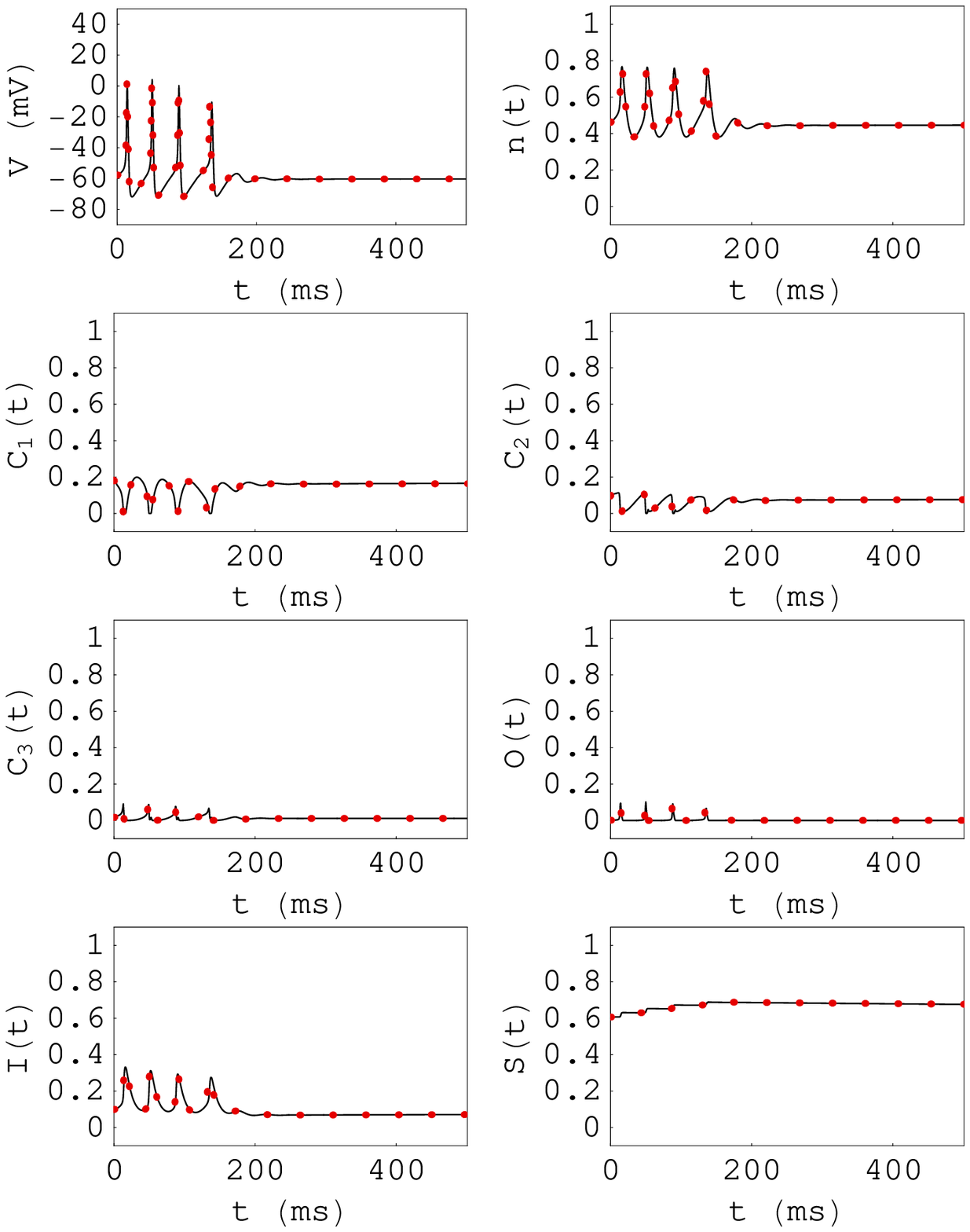}
\caption{The solution of a Na+ channel fifteen state kinetic model, Eqs. (\ref{15c1})  to 
(\ref{15s4})  (solid line) may be approximated by $C_1 = (1 - m)^3 (1 - T)$,
$C_2 = 3m(1 - m)^2 (1 - T)$, $C_3 = 3m^2(1 - m) (1 - T)$, $O = m^3 (1 - T)$, $T = I + S$,
$I = I_2 + I_3 + I_4$ and $S = S_2 + S_3 + S_4$ (dotted line), where $m$,
$T$  and $S$ satisfy Eqs. (\ref{mm3}), (\ref{ratet}) and (\ref{ratesi})  and $n$ and $V$
are determined by Eqs. (\ref{nn}) and (\ref{cur4s}).
 The conditions for the  reduction are that (1) the two stage inactivation process satisfies
 $\beta_{ik} \gg \delta_{ik}$ and   $\gamma_{ik} \gg \alpha_{ik}$,  for each $k$ (see Fig. 11) 
(2) $\alpha_{I1} \gg \rho_1$ and $\sigma_1 \gg \beta_{I1}$ (see Fig. 12) and (3) the transition rates 
between fast inactivated states $I_2$ to $I_4$, and between slow
inactivated states $S_2$ to $S_4$ are an order of magnitude larger  than inactivation and
recovery rates (see Fig. 13). The  increase in the slow inactivation  probability $S$
limits the number of spikes (spike frequency adaptation), and the stationary state of the 
system is stable when the recovery rate $\nu$ for slow inactivation is 
sufficiently small. The rate functions  are $\alpha_{m} = 0.1(V+43.9)/(1 - \exp[-(V+43.9)/10])$,
$\beta_m = 0.108 \exp[-V/19.1]$, $\alpha_{C1} = 3 \alpha_{m }$, $\beta_{C1} = \beta_m$, 
$\alpha_{C2} = 2 \alpha_{m }$, $\beta_{C2} = 2 \beta_m$, $\alpha_{O} = \alpha_{m }$, $\beta_O = 3 \beta_m$,
 $ \alpha_{I1} =  \alpha_{C1}$, $ \beta_{I1} = 0.0135 \beta_{C1}$, $ \alpha_{I2} =  \alpha_{C2}$, 
$ \beta_{I2} = \beta_{C2}$, $ \alpha_{I3} =  \alpha_{O}$, $ \beta_{I3} = \beta_O$,
$\alpha_{ik} = 0.9$, $\gamma_{ik} = 25$, $\beta_{ik} = 2 \exp[-V/10]$, 
$\delta_{i1} = 2.5$,  $\delta_{i2} = \delta_{i3} = \delta_{i4} = 0.0135 \delta_{i1} $, 
$\rho_k =  \alpha_{ik}/(1 + \beta_{ik}/\gamma_{ik})$,  
$\sigma_k = \delta_{ik}/(1 +  \gamma_{ik}/\beta_{ik})$, for $k = 1,4$,
$\mu = 0.047/(1 + \exp[-(V + 17)/10])$, $\nu = 0.00001 \exp[-V/25]$ 
$\alpha_{n} = 0.007(V+58.9)/(1 - \exp[-(V+58.9)/10])$, $\beta_n = 0.038 \exp[-V/80]$ (ms$^{-1}$), and
$\bar{g}_{Na}$= 12 mS/cm$^2$, $\bar{g}_{K}$= 3 mS/cm$^2$, $\bar{g}_{L}$= 0.03 mS/cm$^2$,  $V_{Na}$ = 50 mV,
 $V_{K}$ = -77mV, $V_{L}$ = -54.4 mV, $j = 4$, $C = 1$ $\mu$F/cm$^2$, and $i_{e}$ = 1 $\mu$A/cm$^2$.  
}
\end{center}
\end{figure*}

\begin{figure*}
\begin{center}
\includegraphics[width=0.8 \textwidth]{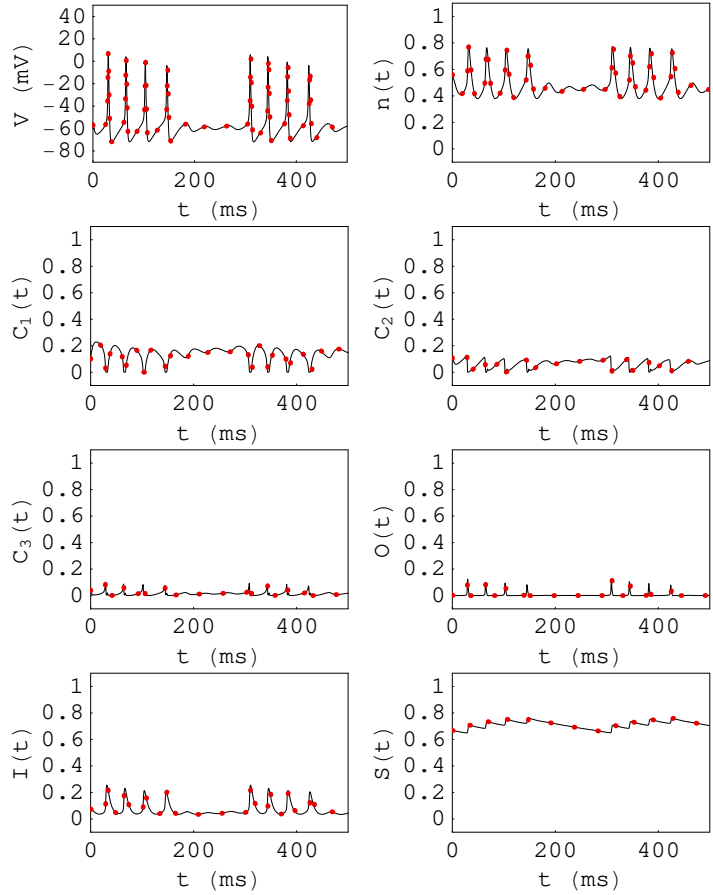}
\caption{The solution of a Na+ channel fifteen state kinetic model, Eqs. (\ref{15c1})  to 
(\ref{15s4})  (solid line) may be approximated  by $C_1 = (1 - m)^3 (1 - T)$,
$C_2 = 3m(1 - m)^2 (1 - T)$, $C_3 = 3m^2(1 - m) (1 - T)$, $O = m^3 (1 - T)$, $T = I + S$,
$I = I_2 + I_3 + I_4$ and $S = S_2 + S_3 + S_4$ (dotted line), where $m$,
$T$  and $S$ satisfy Eqs. (\ref{mm3}), (\ref{ratet}) and (\ref{ratesi}),  and $n$ and $V$
are determined by Eqs. (\ref{nn}) and  (\ref{cur4s}).
 The increase in the slow inactivation probability $S$ terminates the burst
of spikes, and as the slow variable relaxes during the subthreshold oscillation,  the stationary state 
of the subsystem  loses its stability when the recovery rate $\nu$ for slow inactivation is
sufficiently large, and the bursting oscillation resumes. The rate  functions are
$\alpha_{m} = 0.1(V+43.9)/(1 - \exp[-(V+43.9)/10])$, $\beta_m = 0.108 \exp[-V/19.1]$, 
$\alpha_{C1} = 3 \alpha_{m }$, $\beta_{C1} = \beta_m$, $\alpha_{C2} = 2 \alpha_{m }$, 
$\beta_{C2} = 2 \beta_m$, $\alpha_{O} = \alpha_{m}$, $\beta_O = 3 \beta_m$,  $ \alpha_{I1} =  \alpha_{C1}$,
 $ \beta_{I1} = 0.0135 \beta_{C1}$, $ \alpha_{I2} =  \alpha_{C2}$, $ \beta_{I2} = \beta_{C2}$, 
$ \alpha_{I3} =  \alpha_{O}$, $ \beta_{I3} = \beta_O$, 
$\alpha_{ik} = 0.9$, $\gamma_{ik} = 25$, $\beta_{ik} = 2 \exp[-V/10]$, 
$\delta_{i1} = 5.75$,  $\delta_{i2} = \delta_{i3} = \delta_{i4} = 0.0135 \delta_{i1}$, 
$\rho_k =  \alpha_{ik}/(1 + \beta_{ik}/\gamma_{ik})$,  
$\sigma_k = \delta_{ik}/(1 +  \gamma_{ik}/\beta_{ik})$,  for $k = 1,4$,
 $\mu = 0.14/(1 + \exp[-(V + 17)/10])$, $\nu = 0.0001 \exp[-V/25]$, 
$\alpha_{n} = 0.007(V+58.9)/(1 - \exp[-(V+58.9)/10])$, $\beta_n = 0.038 \exp[-V/80]$, (ms$^{-1}$), and
$\bar{g}_{Na}$= 12 mS/cm$^2$, $\bar{g}_{K}$= 3 mS/cm$^2$, $\bar{g}_{L}$= 0.03 mS/cm$^2$,  $V_{Na}$ = 50 mV, 
$V_{K}$ = -77mV, $V_{L}$ = -54.4 mV,  $j = 4$, $C = 1$ $\mu$F/cm$^2$, and $i_{e}$ = 1 $\mu$A/cm$^2$.
}
\end{center}
\end{figure*}

\begin{figure*}
\begin{center}
\includegraphics[width = 0.9 \textwidth]{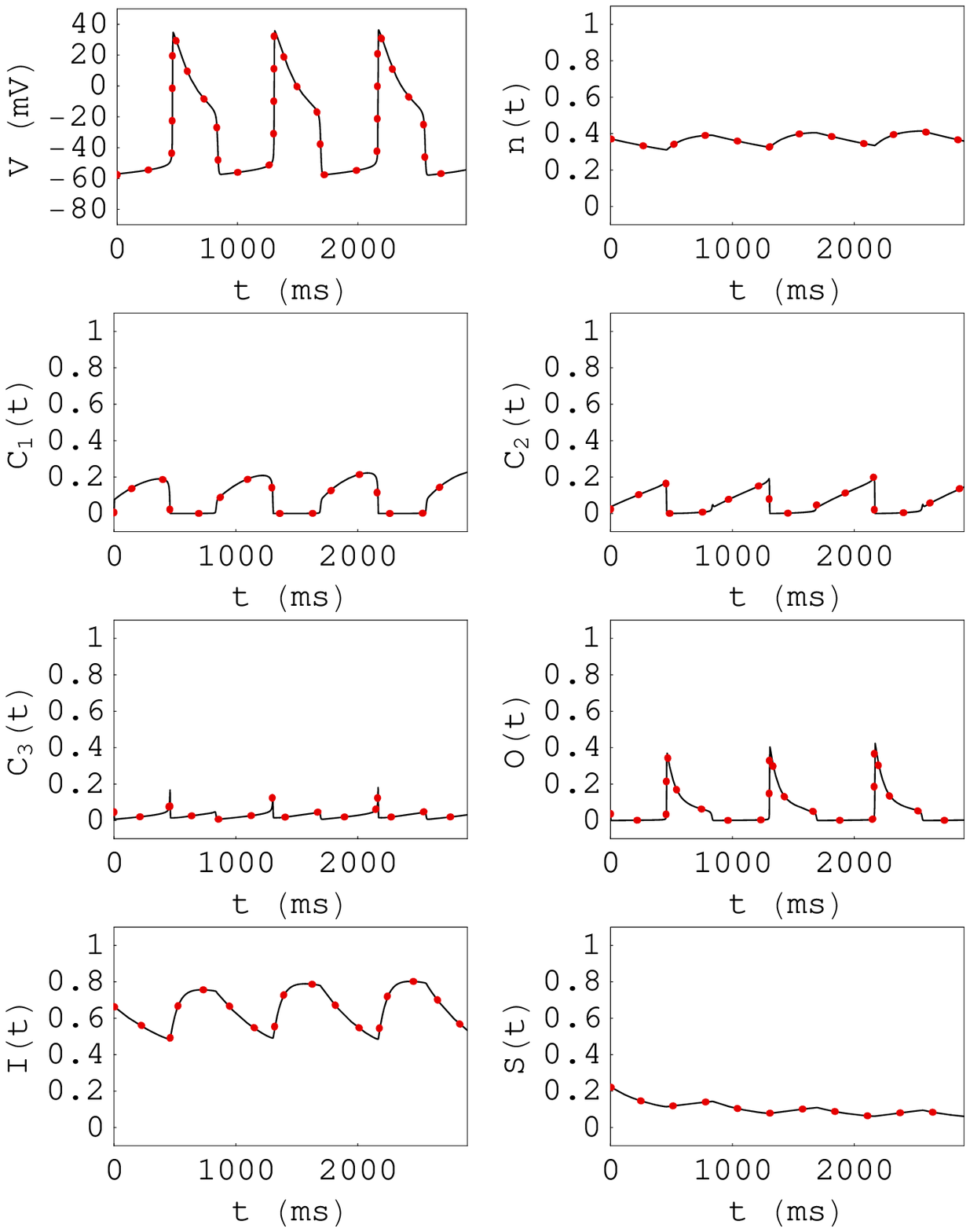}
\caption{The solution of a Na+ channel fifteen state kinetic model, Eqs. (\ref{15c1})  to 
(\ref{15s4})  (solid line) may be approximated by  $C_1 = (1 - m)^3 (1 - T)$,
$C_2 = 3m(1 - m)^2 (1 - T)$, $C_3 = 3m^2(1 - m) (1 - T)$, $O = m^3 (1 - T)$, $T = I + S$,
$I = I_2 + I_3 + I_4$ and $S = S_2 + S_3 + S_4$ (dotted line), where $m$,
$T$  and $S$ satisfy Eqs. (\ref{mm3}), (\ref{ratet}) and (\ref{ratesi}), $n$ and $V$
are determined by Eqs. (\ref{nn}) and  (\ref{cur4s}),  and the rate of recovery from 
inactivation  is sufficiently small to generate a cardiac plateau.  The rate functions are
$\alpha_{m} = 0.1(V+34.3)/(1 - \exp[-(V+34.3)/15])$,
$\beta_m = 4 \exp[-(V+59.3)/25]$, $\alpha_{C1} = 3 \alpha_{m }$, $\beta_{C1} = \beta_m$, 
$\alpha_{C2} = 2 \alpha_{m }$, $\beta_{C2} = 2 \beta_m$, $\alpha_{O} = \alpha_{m }$, $\beta_O = 3 \beta_m$,
 $ \alpha_{I1} =  \alpha_{C1}$, $ \beta_{I1} = 0.0135 \beta_{C1}$, $ \alpha_{I2} = 10 \alpha_{C2}$, 
$ \beta_{I2} = \beta_{C2}$, $ \alpha_{I3} =  \alpha_{O}$, $ \beta_{I3} = \beta_O$,
$\alpha_{ik} = 0.012$, $\gamma_{ik} = 25$, $\beta_{ik} = 4.2 \exp[-2.3(V - 31.9)/25]$, 
$\delta_{i1} = 0.1$,  $\delta_{i2} = 0.0135 \delta_{i1}$, $\delta_{i3} = \delta_{i4} = 0.00135 \delta_{i1}$ 
$\rho_k =  \alpha_{ik}/(1 + \beta_{ik}/\gamma_{ik})$,  
$\sigma_k = \delta_{ik}/(1 +  \gamma_{ik}/\beta_{ik})$, for $k = 1,4$ (ms$^{-1}$),
$\mu = 0.06 \exp[0.1(V+11)/25]$ (s$^{-1}$), $\nu = 0.108 \exp[-1.95(V+11)/25]$ (s$^{-1}$), 
$\alpha_{n} = 0.015(V+25)/(1 - \exp[-(V+25)/10])$, $\beta_n = 0.5 \exp[-(V+65)/80]$ (s$^{-1}$), and
 $\bar{g}_{Na}$= 36 mS/cm$^2$, $\bar{g}_{K}$= 3 mS/cm$^2$, $\bar{g}_{L}$= 2 mS/cm$^2$,  $V_{Na}$ = 55 mV,
 $V_{K}$ = -80mV, $V_{L}$ = -58.5 mV,  $j = 1$, $C = 12$ $\mu$F/cm$^2$,  and $i_{e}$ = 27 $\mu$A/cm$^2$.
}
\end{center}
\end{figure*}

\begin{figure*}
\begin{center}
\includegraphics[width=0.9 \textwidth]{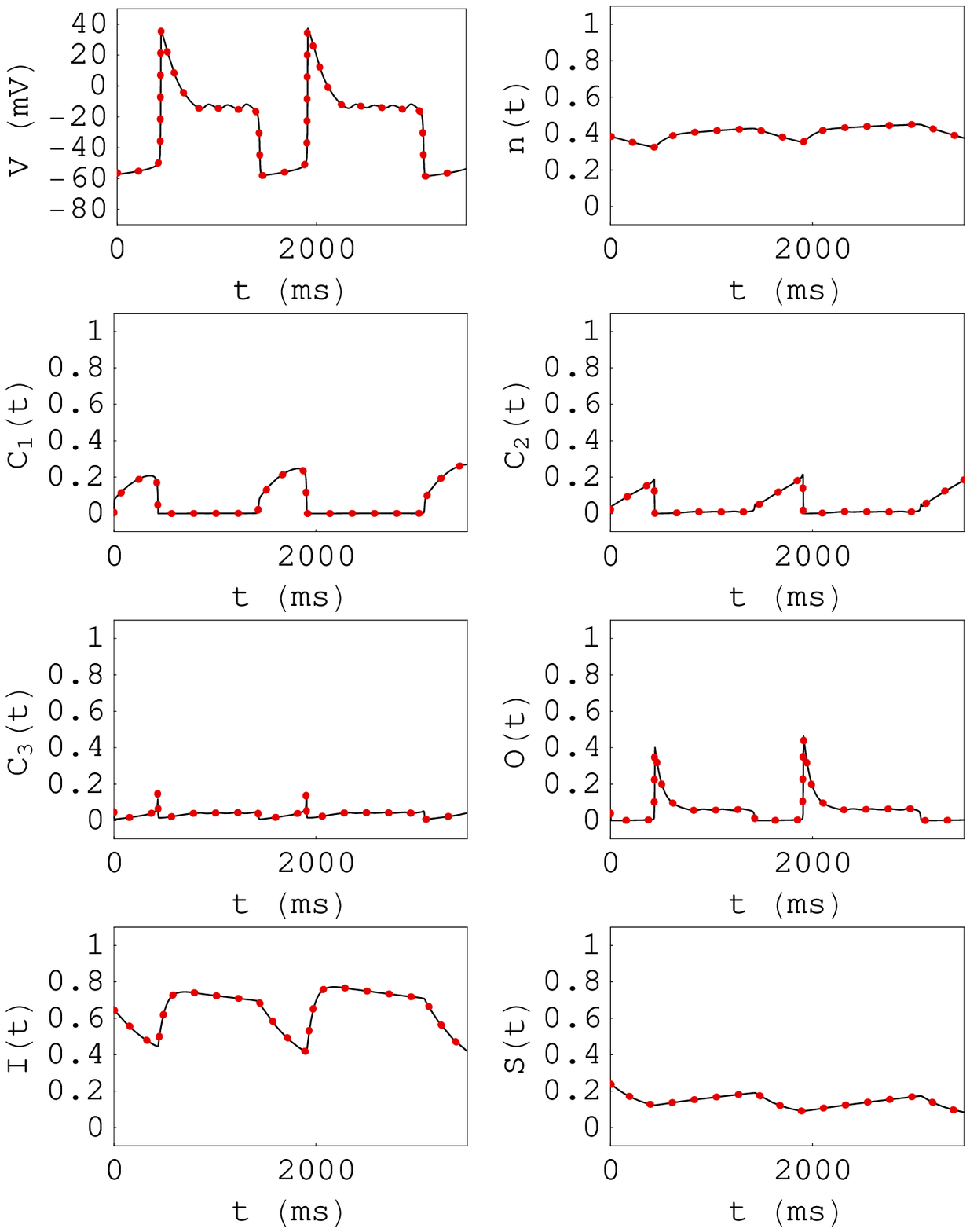}
\caption{The solution of a Na+ channel fifteen state kinetic model, Eqs. (\ref{15c1})  to 
(\ref{15s4})   (solid line) may be approximated  by $C_1 = (1 - m)^3 (1 - T)$,
$C_2 = 3m(1 - m)^2 (1 - T)$, $C_3 = 3m^2(1 - m) (1 - T)$, $O = m^3 (1 - T)$, $T = I + S$,
$I = I_2 + I_3 + I_4$ and $S = S_2 + S_3 + S_4$ (dotted line), where $m$,
$T$  and $S$ satisfy Eqs. (\ref{mm3}), (\ref{ratet}) and (\ref{ratesi}), $n$ and $V$
are determined by Eqs. (\ref{nn}) and (\ref{cur4s}), and the rate of recovery   from inactivation
$\sigma_1$ is increased to generate a cardiac action potential with a plateau oscillation.  
The rate functions are $\alpha_{m} = 0.1(V+34.3)/(1 - \exp[-(V+34.3)/15])$,
$\beta_m = 4 \exp[-(V+59.3)/25]$, $\alpha_{C1} = 3 \alpha_{m }$, $\beta_{C1} = \beta_m$, 
$\alpha_{C2} = 2 \alpha_{m }$, $\beta_{C2} = 2 \beta_m$, $\alpha_{O} = \alpha_{m }$, $\beta_O = 3 \beta_m$,
$ \alpha_{I1} =  \alpha_{C1}$, $ \beta_{I1} = 0.0135 \beta_{C1}$, $ \alpha_{I2} = 10 \alpha_{C2}$, 
$ \beta_{I2} = \beta_{C2}$, $ \alpha_{I3} =  \alpha_{O}$, $ \beta_{I3} = \beta_O$,
$\alpha_{ik} = 0.012$, $\gamma_{ik} = 25$, $\beta_{ik} = 4.2 \exp[-2.3(V - 31.9)/25]$, 
$\delta_{i1} = 0.12$,  $\delta_{i2} = 0.0135 \delta_{i1}$, $\delta_{i3} = \delta_{i4} = 0.00135 \delta_{i1}$  
$\rho_k =  \alpha_{ik}/(1 + \beta_{ik}/\gamma_{ik})$,  
$\sigma_k = \delta_{ik}/(1 +  \gamma_{ik} /\beta_{ik})$, for $k = 1,4$ (ms$^{-1}$),
$\mu = 0.06 \exp[0.1(V+11)/25]$ (s$^{-1}$), $\nu = 0.108 \exp[-1.95(V+11)/25]$ (s$^{-1}$), 
$\alpha_{n} = 0.015(V+25)/(1 - \exp[-(V+25)/10])$, $\beta_n = 0.5 \exp[-(V+65)/80]$ (s$^{-1}$), and 
and $\bar{g}_{Na}$ = 36 mS/cm$^2$, $\bar{g}_{K}$= 3 mS/cm$^2$, $\bar{g}_{L}$= 2 mS/cm$^2$,  $V_{Na}$ = 55 mV,
$V_{K}$ = -80mV, $V_{L}$ = -58.5 mV,  $j = 1$,$C = 12$ $\mu$F/cm$^2$, and $i_{e}$ = 27 $\mu$A/cm$^2$.
}
\end{center}
\end{figure*}

\end{document}